\documentclass[a4paper,11pt,twoside]{scrreprt} 
\usepackage[left=2.5cm,right=2cm,top=2cm,bottom=2cm]{geometry}

\usepackage[utf8]{inputenc}
\usepackage[T1]{fontenc}

\usepackage{amsfonts}
\usepackage{amsthm}
\usepackage{amssymb}

\usepackage{tikz}
\usetikzlibrary{shapes.geometric, arrows}

\tikzstyle{startstop} = [rectangle, rounded corners, minimum width=3cm, minimum height=1cm,text centered, draw=black, fill=red!30]
\tikzstyle{io} = [trapezium, trapezium left angle=70, trapezium right angle=110, minimum width=3cm, minimum height=1cm, text centered, draw=black, fill=blue!30]
\tikzstyle{process} = [rectangle, minimum width=3cm, minimum height=1cm, text centered, draw=black, fill=orange!30]
\tikzstyle{decision} = [diamond, minimum width=3cm, minimum height=1cm, text centered, draw=black, fill=green!30]
\tikzstyle{arrow} = [thick,->,>=stealth]

\usepackage{libertine}

\usepackage{float}

\usepackage{listings}

\usepackage[square, comma, sort&compress]{natbib}

\usepackage[hyphens]{url}

\usepackage{proof}

\usepackage[normalem]{ulem}

\usepackage[lined,ruled,commentsnumbered]{algorithm2e}

\usepackage[toc]{glossaries}
\makeglossaries
\include{glossary}
\let\Oldgls\gls \renewcommand{\gls}[1]{\dashuline{\Oldgls{#1}}}

\theoremstyle{definition}

\date{\today}

\begin{document}


\newcommand{\reporttitle}{Implementing a teleo-reactive programming system}
\newcommand{\reportauthor}{Robert Webb}
\newcommand{\supervisor}{Dr Anthony Field}
\newcommand{\degreetype}{Advanced Computing}

\begin{titlepage}

\newcommand{\HRule}{\rule{\linewidth}{0.5mm}} 


\includegraphics[width = 4cm]{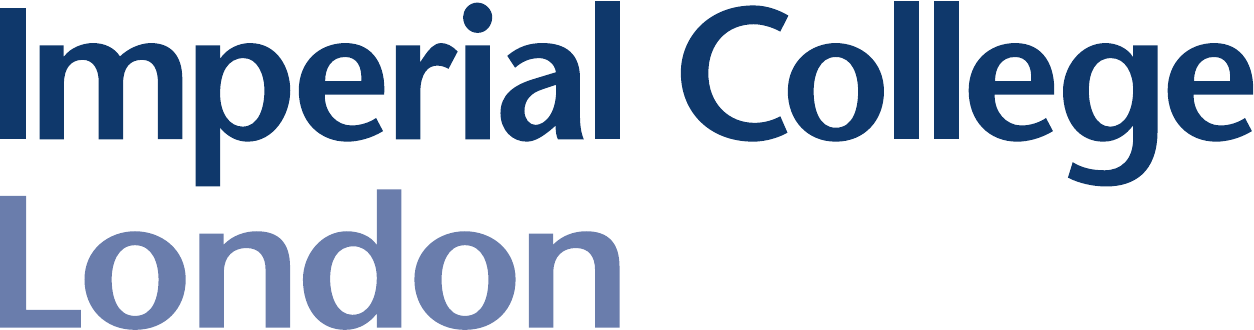}\\[0.5cm] 

\center 


\textsc{\Large Imperial College London}\\[0.5cm] 
\textsc{\large Department of Computing}\\[0.5cm] 


\HRule \\[0.4cm]
{ \huge \bfseries \reporttitle}\\ 
\HRule \\[1.5cm]
 

\begin{minipage}{0.4\textwidth}
\begin{flushleft} \large
\emph{Author:}\\
\reportauthor 
\end{flushleft}
\end{minipage}
~
\begin{minipage}{0.4\textwidth}
\begin{flushright} \large
\emph{Supervisor:} \\
\supervisor 
\end{flushright}
\end{minipage}\\[4cm]

\vfill 
Submitted in partial fulfillment of the requirements for the MSc degree in
\degreetype~of Imperial College London\\[0.5cm]

\makeatletter
\@date 
\makeatother

\end{titlepage}

\addchap*{Acknowledgements}

I would like to express my gratitude to Dr Anthony Field for his support and guidance throughout my project.\\

I would also like to thank Dr Keith Clark and Dr Peter Robinson for their support with the Qulog software and discussion in person and via email.\\

Finally, I thank my family, my friends and all my professors at Imperial College.\\

\begin{abstract}
This thesis explores the teleo-reactive programming paradigm for controlling autonomous agents, such as robots. Teleo-reactive programming provides a robust, opportunistic method for goal-directed programming that continuously reacts to the sensed environment. In particular, the \textsc{TR} and \textsc{TeleoR} systems are investigated. They influence the design of a teleo-reactive system programming in Python, for controlling autonomous agents via the Pedro communications architecture. To demonstrate the system, it is used as a controller in a simple game.
\end{abstract}

\tableofcontents

\chapter{Introduction}
Teleo-reactive programming is a programming paradigm for autonomous agents (such as robots) that offers a way to deal with the unpredictable nature of the real world, as well as the challenge of connecting continuously-sensed inputs (percepts, sensed data) to outputs (actions). It incorporates ideas from control theory such as continuous feedback but it also incorporates features from computer science such as procedures and variable instantiation\citep{Nilsson1994}.\\

\section{Objectives}
One of the reasons why robotics is so challenging is that the real world is unpredictable. If a robot sets out to achieve some action, unexpected things might happen that ``set back'' the robot's progress or cause the robot to fortuitously skip a few steps in its algorithm. Another challenge is that the world changes continuously, while computers and computer programs operate in terms of discrete time-steps. Teleo-reactive programming both of these problems by allowing robust (able to recover from setbacks) and opportunistic (able to take advantages of fortuitous changes in the environment) computer programs to be developed, that also continuously react to their sensed environment.\\

This report aims to describe in detail the seminal teleo-reactive system \textsc{TR}\citep{Nilsson1994,Nilsson2001} and the later system \textsc{TeleoR}\citep{Clark2014,Clarka}, which adds additional features to the original system. It also describes the \textbf{Pedro} communications protocol, which is used by \textsc{TeleoR} for inter-agent communication.\\

This report also describes a teleo-reactive system developed for this thesis project, which consists of an interpreter of a \textsc{TR}-like language which can communicate a simulation via Pedro, implemented in Python. The syntax and semantics (how the language is evaluated) will be explained in detail.\\

The teleo-reactive system will be demonstrated controlling a simple demonstration program. Sample teleo-reactive programs and explanations of their meaning are given. The report will also discuss whether teleo-reactive solves the problem of robot control well, what alternatives to \textsc{TR} and \textsc{TeleoR} exist, the alternatives to teleo-reactive programming and the miscellaneous practical considerations made when developing this project.

\section{Overview}
The structure of the report is as follows:\\
\textbf{Chapter \ref{ch:background}} provides background knowledge, describing Nilsson's \textsc{TR} system and Clark \& Robinson's \textsc{TeleoR} system.\\
\textbf{Chapter \ref{ch:language}} describes the language of the system that was developed at a high level, covering the syntax.\\
\textbf{Chapter \ref{ch:implementation}} describes in more detail how the system works, the algorithms involved, the type system and what programs are (in)valid.\\
\textbf{Chapter \ref{ch:evaluation}} presents a demonstration of the teleo-reactive system, gives an evaluation of the success of the system, the alternative teleo-reactive systems (other than \textsc{TR} and \textsc{TeleoR}) that have been developed, alternatives to teleo-reactive programming and practical decisions made during the project.\\
\textbf{Chapter \ref{ch:conclusion}} sums up the project and suggests future related areas of research.

\chapter{Background}\label{ch:background}
\section{Teleo-reactive programming}
\textbf{Teleo-reactive programming} involves creating programs whose actions are all directed towards achieving goals, in continuous response to the state of the environment.  It offers a way to specify robust, opportunistic and goal-directed robot behaviour \citep{Clark2014}. This means that it recovers from setbacks and will skip unnecessary actions if it can\citep{Clarka}. \\

This section will explain teleo-reactive programming, by describing Nilsson's \textsc{TR} language\citep{Nilsson1994, Nilsson2001}. It will then describe \textsc{TeleoR}, a language by Clark and Robinson that extends the semantics of \textsc{TR} while adding performance optimisations and facilities for multi-agent programming. The two languages are intended as mid-level languages, they act as an interface between the senses and the actions performed by the agent\citep{Clarka}.\\

\subsection{Continuous evaluation}
Designing autonomous agents is difficult because they must operate in a constantly changing environment which can be sensed only imperfectly and only affected with uncertain results. However, autonomous agents have been developed in other domains that do function effectively in the real world for long periods of time. For example, governors that control the speed of steam engines, thermostats and complex guidance systems. One thing that these systems have in common is that they continuously respond to their environment\citep{Nilsson1992}.\\

Nilsson proposes teleo-reactive programming as a paradigm that allows programs to be written that continuously respond to their inputs in a similar way that the output of an electronic circuit continuously responds to its input signals, but retains useful concepts from computer science such as hierarchical organisation of programs, parameters, routines and recursion\citep{Nilsson1992}. He refers to computer programs that continuously respond to their inputs as having ``circuit semantics''. Nilsson also intended for teleo-reactive programs to be responsive to the agent's stored model of the environment, as well as directly sensed data\citep{Nilsson1994}.\\

\newpage
\section{\textsc{TR}}
The \textsc{TR} language is Nilsson's implementation of teleo-reactive programming\citep{Nilsson2001}. This chapter will describe the design and features of the \textsc{TR} language, then explore the later \textsc{TeleoR} language which adds additional features to \textsc{TR}.

\subsection{Triple-tower architecture}
At the most abstract level, Nilsson splits the problem of decision making in autonomous agents into three tasks, which are performed by the \textbf{triple-tower architecture}\citep{Nilsson2001}. The three parts of the system that perform these three tasks are called the ``towers'' of the architecture, because they work at multiple levels of abstraction, that incrementally build on lower layers. The three towers are:
\begin{itemize}
\item the perception tower - deducing further truths about the world (rules);
\item the model tower - maintaining the agent's knowledge about the world (predicates and truth maintenance system);
\item the action tower - deciding what course of action to take (action routines);
\end{itemize}

\begin{figure}[ht!]
\centering
\includegraphics[width=90mm]{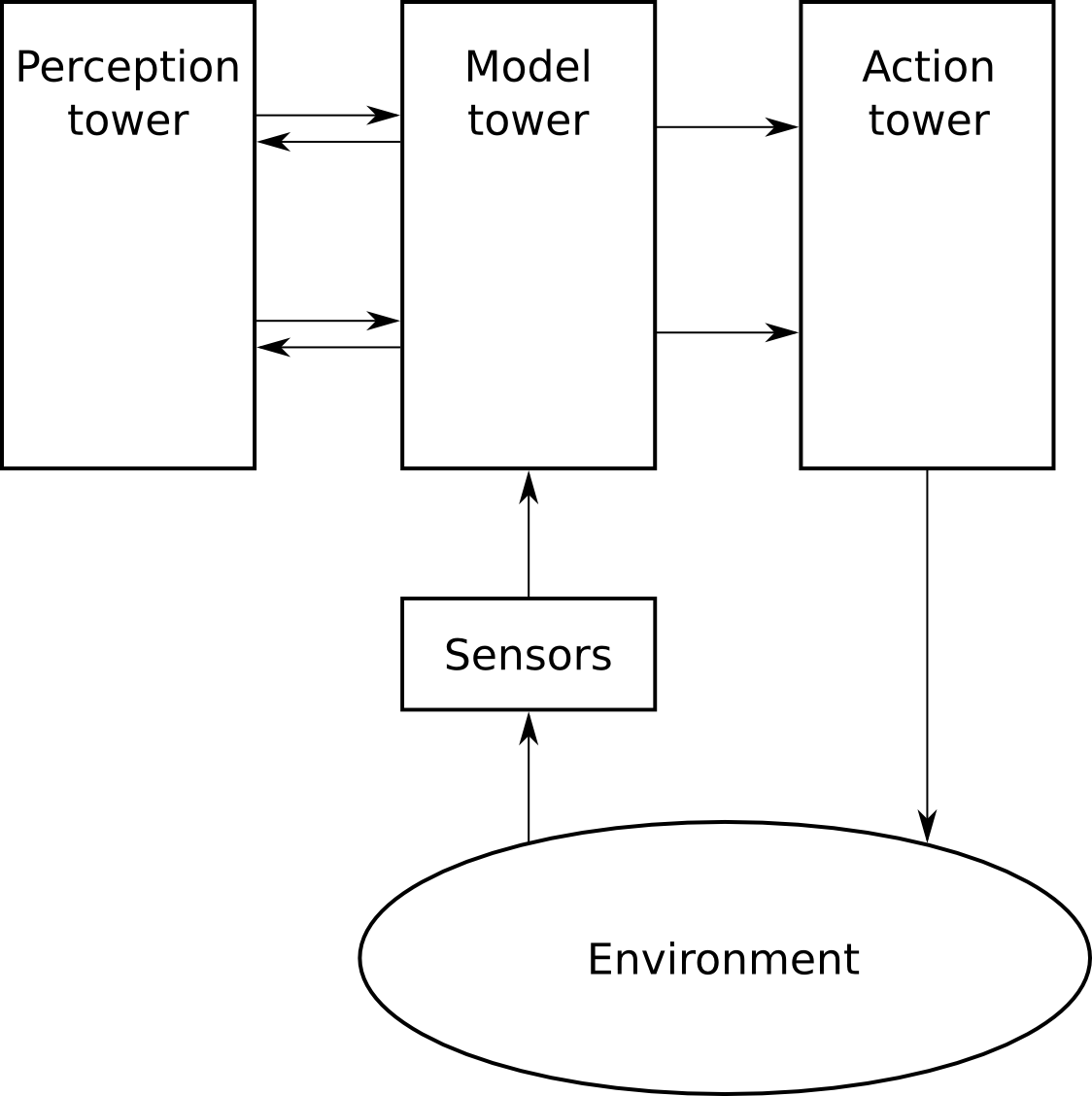}
\caption{Diagram of the triple-tower architecture \label{tripletowerdiagram}}
\end{figure}

\subsubsection{Perception tower}
The first tower is the \textbf{perception tower}. It consists of logical rules that are used to deduce new facts from existing knowledge. These can be expressed using a logic programming language. Each rule can be defined in terms of other rules, hence the ``tower'' nature of the perception tower. The perception tower creates higher-abstraction percepts from lower-abstraction ones\citep{Nilsson2001}.\\

For example, consider a scenario with a percept \texttt{on(X,Y)}, objects \texttt{block(X)} (uniquely numbered, for \(X \in \mathbb{N}\)), the object \texttt{table} and the object \texttt{nothing}. \texttt{on(X,Y)} means that the object \texttt{X} is on top of the object \texttt{Y}. From this basic percept, a new predicate \texttt{sorted\_stack(X)} can be defined which states that the block \texttt{X} is at the top of a stack of blocks, where the blocks are sorted with \texttt{block(1)} at the top and the highest numbered block at the bottom. This could be defined as follows:\\
\begin{verbatim}
sorted_stack(table).

sorted_stack(block(1)) :- on(empty, block(1)),
                          on(block(1), X),
                          sorted_stack(X).

sorted_stack(block(N)) :- on(block(N),
                          block(M)),
                          N >= M,
                          sorted_stack(block(M)).
\end{verbatim}
This new predicate \texttt{sorted\_stack(X)} can now be queried by another part of the program. This example uses Prolog expressions.

\subsubsection{Model tower}
The \textbf{model tower} consists of a truth-maintenance system that invokes the rules in the perception tower on the percepts sensed by the agent. The goal of the TMS is to keep the agent's knowledge continuously faithful to changes in the environment\citep{Nilsson2001}.

\subsubsection{Action tower}
The \textbf{action tower} specifies what actions the robot should perform, based on the knowledge maintained by the model tower. It should be possible to define actions in terms of other (sub-)actions, so that complex behaviours can be defined in terms of simpler ones. For example, picking up a box involves moving an arm, opening and closing a gripper, etc. \\

The \textsc{TR} language was designed by Nilsson to perform the role of the action tower. Given facts about the world (percepts) and the robot's knowledge (stored beliefs and predicates inferred from the beliesf and percepts), the robot can autonomously perform actions based on a series of rules\citep{Nilsson2001}. The next few sections will explain what the \textsc{TR} language is and how it can be used. \\

\subsection{\textsc{TR} sequences}
The distinctive feature of teleo-reactive programming is the \textbf{\textsc{TR} sequence}, which is an ordered sequence of rules of the form ``if some conditions are satisfied, then perform these actions''\citep{Nilsson1994}. The conditions, or \textbf{guards} are defined in terms of \textbf{percepts} and \textbf{beliefs}. Percepts are information that has been sensed by the agent and beliefs are facts remembered by the agent. An action can be either be a tuple of primitive actions to be performed concurrently (e.g. move forward, turn left, look up) or a call to initiate another \textsc{TR} sequence\citep{Clarka}. \\

The actions currently being performed by the agent are defined as the actions associated with the first satisfied condition in the \textsc{TR} sequence. If no condition of any of the rules can be satisfied, an error occurs, so the last rule is sometimes written so that it always fires (being a `none of the above' condition)\citep{Clarka}.

\begin{verbatim}
K_1 ~> A_1
K_2 ~> A_2
...
K_n ~> A_n
\end{verbatim}

In order for a \textsc{TR} sequence to achieve a goal, the rules are written in a way that the topmost rule describes the goal state and every other rule brings the agent closer to satisfying the guards of the rules above them. A \textsc{TR} sequence with this feature is referred to as having the \textbf{regression property}. If a \textsc{TR} sequence has the regression property and a guard can always be satisfied (e.g. if the bottommost rule always applies), then it is a \textbf{universal program}\citep{Clarka}.

\subsubsection{Example}
To give a simple example of a \textsc{TR} sequence, imagine a 2D world where there is a robot and a light. The goal of this procedure is for the robot to turn to face the light.
\begin{itemize}
\item The first rule defines the condition that the procedure is supposed to achieve - the robot is facing the light, in which case, do nothing.
\item The second rule says that if the robot sees the light to the left of it, then turn left.
\item The third rule says that if the robot sees the light to the right of it, then turn right.
\item The fourth rule is fired if the robot does not see anything - it continually turns left until something comes into view that triggers the upper rules. 
\end{itemize}
\begin{verbatim}
is_facing(light) ~> ()
see(light, left) ~> turn(left)
see(light, right) ~> turn(right)
() ~> turn(left)
\end{verbatim}

\subsection{\textsc{TR} procedures}
Goals can be split into sub-goals, sub-sub-goals and so on, so to reflect this teleo-reactive programs can also be written in a hierarchical and structured way, using \textsc{TR} procedures. A \textbf{\textsc{TR} procedure} is a \textsc{TR} sequence that can be called by another \textsc{TR} sequence. Just like a procedure in a structured imperative language, it has a name and takes a tuple of parameters as input. \textsc{TR} procedures can be written that achieve sub-goals, that are then called by procedures to achieve higher goals \citep{Clark2014}. For example, a program that tells an agent to pick up a box with a robotic arm could be made up of:

\begin{itemize}
\item a procedure that tells the agent to face the box
\item a procedure that tells the agent to move forwards to the box
\item a procedure that tells the agent to pick up the box that is in front of it
\end{itemize}

The syntax of a \textsc{TR} procedure is as follows:
\begin{verbatim}
procedure_name(param_1, param_2, ... , param_k){
  K_1 ~> A_1
  K_2 ~> A_2
  ...
  K_n ~> A_n
}
\end{verbatim}
\texttt{procedure\_name} is the name of the procedure, \texttt{param\_1, ... , param\_k} are names of the parameters passed to the procedure.

Guards in \textsc{TR} procedures can be partially instantiated, which means that some (or none) of the variables in the guard are from the parameters of the procedure, some are constant terms and some become instantiated when the rule is satisfied\citep{Clarka}. While not all variables on the left hand side of the rule have to be instantiated, all variables on the right hand side must be instantiated, otherwise an error is thrown. The ability to write rules in this way means that more general rules can be written, making the program code shorter and more readable.

\subsubsection{Example}
The procedure from the previous example could be rewritten as a \textsc{TR} procedure:
\begin{verbatim}
face_thing(Thing){
  is_facing(Thing) ~> ()
  see(Thing, Dir) ~> turn(Dir)
  () ~> turn(left)
}
\end{verbatim}

Calling this procedure as \texttt{face\_thing(light)} causes the variable \texttt{Thing} to be instantiated to \texttt{light}. Rewriting the original sequence as a procedure means that the robot can be told to face many different things, using the same code. The second and third lines of the original sequence have also been combined into a single rule. If the robot sees the light to the left hand side, then \texttt{Dir} is instantiated as \texttt{left} and the action \texttt{turn(left)} is performed. Likewise, if the robot sees the light to the right hand side, then \texttt{Dir} is instantiated as \texttt{right} and the action \texttt{turn(right)} is performed.
\newpage
\subsubsection{\textsc{TR} algorithm}
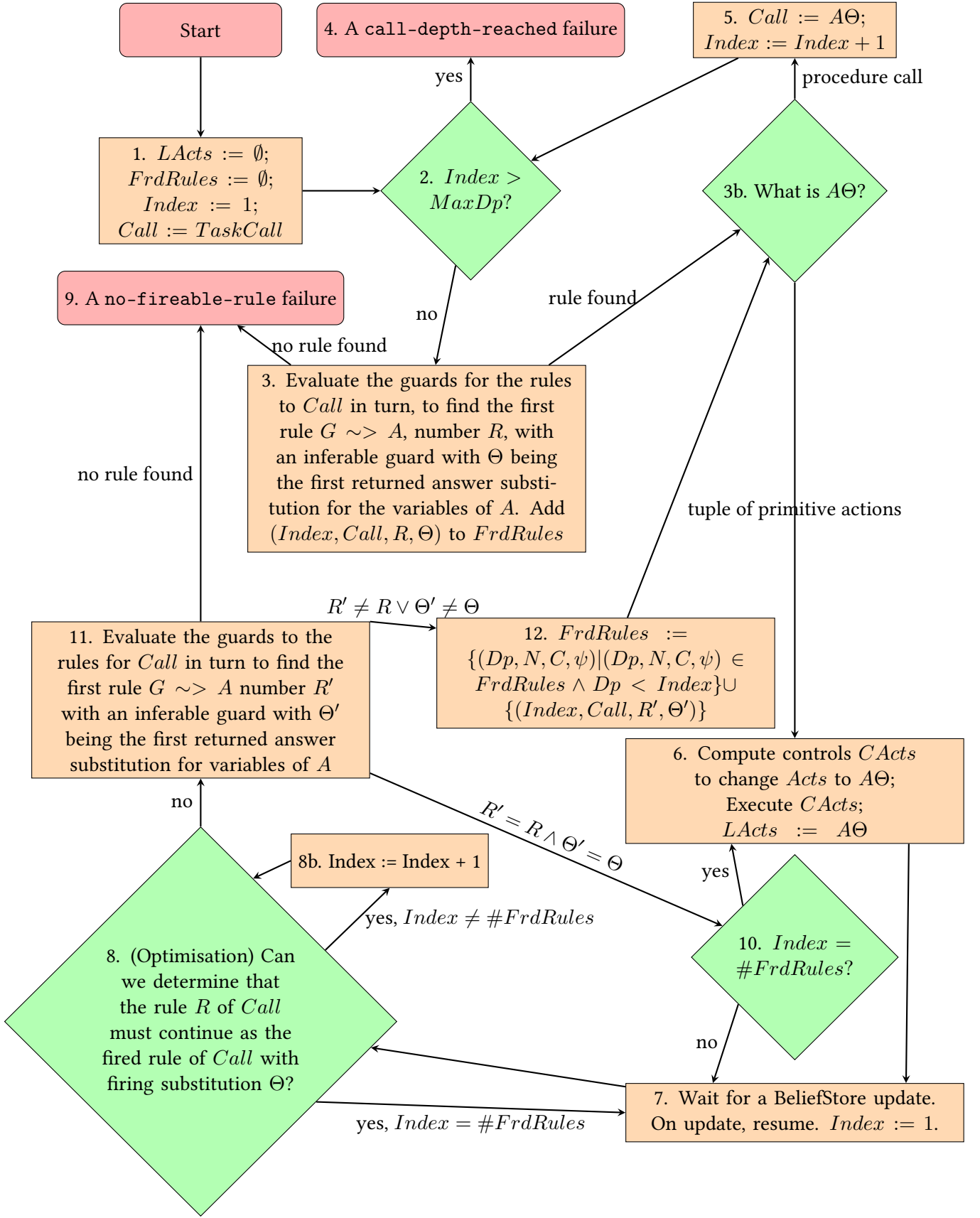
\begin{figure}[H]
\begin{tikzpicture}[node distance=2cm]

\node (start) [startstop] {Start};

\node (step1) [process, below of=start, minimum width=3.5cm, text centered, text width=3.5cm, yshift=-1cm] {1. \(LActs := \emptyset;\) \\ \(FrdRules := \emptyset;\) \\ \(Index := 1;\) \\ \(Call := TaskCall\)};

\node (step2) [decision, right of=step1, xshift=3cm, text width=2cm] {2. \(Index > MaxDp?\)};

\node (step3) [process, below of=step2, minimum width=3cm, minimum height=1cm, text centered, text width=6cm, yshift=-3cm, xshift=-1cm] {3. Evaluate the guards for the rules to \(Call\) in turn, to find the first rule \(G \sim > A\), number \(R\), with an inferable guard with \(\Theta\) being the first returned answer substitution for the variables of \(A\). Add \((Index, Call, R, \Theta)\) to \(FrdRules\)};

\node (step3b) [decision, right of=step2, xshift=4cm] {3b. What is \(A\Theta\)?};

\node (step4) [startstop, right of=start, xshift=3cm] {4. A \texttt{call-depth-reached} failure};

\node (step5) [process, right of=step4, minimum width=3.5cm, text centered, text width=3.5cm, xshift=4cm] {5. \(Call := A\Theta;\) \\ \(Index := Index + 1\)};

\node (step6) [process, below of=step3b, minimum width=3cm, minimum height=1cm, text centered, text width=6cm, yshift=-9.2cm] {6. Compute controls \(CActs\) to change \(Acts\) to \(A\Theta\); \\ Execute \(CActs\); \\ \(LActs := A\Theta\)};

\node (step7) [process, below of=step6, minimum width=3cm, minimum height=1cm, text centered, text width=6cm, yshift=-4cm] {7. Wait for a BeliefStore update. \\ On update, resume. \(Index := 1\).};

\node (step9) [startstop, below of=step1] {9. A \texttt{no-fireable-rule} failure};

\node (step11) [process, below of=step1, minimum width=3cm, minimum height=1cm, text centered, text width=6cm, yshift=-7.5cm] {11. Evaluate the guards to the rules for \(Call\) in turn to find the first rule \(G \sim > A\) number \(R'\) with an inferable guard with \(\Theta'\) being the first returned answer substitution for variables of \(A\)};

\node (step8) [decision, below of=step11, minimum width=2cm, minimum height=1cm, text centered, text width=4cm, yshift=-4cm] {8. (Optimisation) Can we determine that the rule \(R\) of \(Call\) \\ must continue as the fired rule of \(Call\) with firing substitution \(\Theta\)?};

\node (step12) [process, below of=step3, minimum width=3cm, minimum height=1cm, text centered, text width=6cm, yshift=-2cm, xshift=3.5cm] {12. \(FrdRules := \{(Dp, N, C, \psi) | (Dp, N, C, \psi) \in FrdRules \land Dp < Index\} \cup \) \\ \( \{(Index, Call, R', \Theta')\}\)};

\node (step10) [decision, below of=step12, minimum width=2cm, minimum height=1cm, text centered, text width=2.5cm, yshift=-3.3cm, xshift=3.5cm] {10. \(Index = \#FrdRules?\)};

\node (step13) [process, below of=step11, minimum width=3cm, text centered, xshift=3.5cm, yshift=-1cm] {8b. Index := Index + 1};

\draw [arrow] (start) -- (step1);

\draw [arrow] (step1) -- (step2);

\draw [arrow] (step2) -- node[anchor=east] {no} (step3);

\draw [arrow] (step2) -- node[anchor=east] {yes} (step4);

\draw [arrow] (step3b) -- node[anchor=west] {procedure call} (step5);

\draw [arrow] (step3b) -- node[anchor=center] {tuple of primitive actions} (step6);

\draw [arrow] (step3) -- node[anchor=west] {no rule found} (step9);

\draw [arrow] (step3) -- node[anchor=east] {rule found} (step3b);

\draw [arrow] (step5) -- (step2);

\draw [arrow] (step6.335) -- (step7.15);

\draw [arrow] (step7) -- (step8);

\draw [arrow] (step8.325) -- node[anchor=north] {yes, \(Index = \#FrdRules\)} (step7.180);

\draw [arrow] (step8) -- node[anchor=east] {no} (step11);

\draw [arrow] (step10.135) -- node[anchor=east] {yes} (step6.220);

\draw [arrow] (step10.220) -- node[anchor=east] {no} (step7.160);

\draw [arrow] (step11.25) -- node[anchor=south] {\(R' \neq R \lor \Theta' \neq \Theta\)}  (step12.165);
\draw [arrow] (step11) -- node[anchor=south, sloped] {\(R' = R \land \Theta' = \Theta\)} (step10);

\draw [arrow] (step11) -- node[anchor=east] {no rule found} (step9);

\draw [arrow] (step12) -- (step3b);

\draw [arrow] (step13.180) -- (step8.70);
\draw [arrow] (step8.30) -- node[anchor=west] {yes, \(Index \neq \#FrdRules\)} (step13.270);
\end{tikzpicture}
\caption{Flowchart depicting the \textsc{TR} algorithm\label{fig:flowchart}}
\end{figure}

The flowchart on the previous page (Figure \ref{fig:flowchart}) describes an informal algorithm\citep{Clark2014} for executing \textsc{TR} programs. It describes the operation of a \textsc{TR} program that has been called with the call \(TaskCall\). \(FrdRules\) is the set of indexed active procedure calls. Each element of \(FrdRules\) is a tuple of the form \((Dp, Call, R, \Theta)\) where \(Dp - 1\) is the number of intermediary procedure calls between \(Call\) and \(TaskCall\), \(R\) is the number of the partially instantiated rules of the procedure for \(Call\) that was last fired and \(\Theta\) is the set of generated bindings for all the variables of the action of that rule. \(Dp\) is the index of the tuple. \(Acts\) is the last tuple of determined actions for \(TaskCall\), initialised to \(()\).\\

Step 1 initialises the state of the algorithm, which consists of \(LActs\), \(FrdRules\), \(Index\) and \(Call\).\\

Step 2 is the first step of the execution loop of the program, it checks that the maximum call depth has not been exceeded. The maximum depth is defined by the programmer with the parameter \(MaxDp\). Step 3 finds the first rule in the current call. If no rule is fired, the algorithm goes to step 9 and fails. If a rule is found, the algorithm goes to step 3b. If the rule's action (with variables instantiated) is a procedure call, then call the procedure and go back to step 2. If it is a tuple of primitive actions, go to step 6 and compute the controls and execute them, then go to step 7.\\

At step 7, wait for a BeliefStore update (i.e. a modification to the beliefs and percepts), then set \(Index\) to \(1\). Step 8 and 8b consist of a loop that checks every previously fired rule from the last call to see if it must continue to fire. This is done by keeping track of which predicates must be true or false to satisfy the guard conditions for each rule firing. If every previously fired rule has been found to continue, go back to step 7. Otherwise, go to step 11.\\

At step 11, re-evaluate the guards in the same way as in step 3 to find the rule that must now fire. If the rule and variable substitution are the same, go to step 10. If either of them are different, update \(FrdRules\) so that the old rule at the current level in the call stack is replaced by the new rule, then go to step 3b. If no rule can be found, go to step 9 (failure)\citep{Clarka}.\\

\subsection{Continuation of firings}\label{subsec:continuation}
From the rule definitions, it is possible to know which predicates would have been queried to find the first fireable rule. With this information, it is possible to determine the conditions (i.e. which predicates must be inferable or not inferable) under which it would continue to fire.\\

There are only two ways in which a currently firing rule can be interrupted:
\begin{itemize}
\item The guard of a rule above the previously fired rule becomes inferable;
\item The conditions for the previously fired rule become inferable for a different instantiation of variables.
\end{itemize}

Therefore, unless one of these two conditions holds, there is no need to re-evaluate the guards of the \textsc{TR} procedure. This can be expressed in terms of the ``local dependent predicates'' for a given rule, inside a \textsc{TR} procedure.  This is a list of predicates, that if they were to become inferable / not inferable, would cause a given rule to no longer continue to fire. This is defined as a list of functors (names of predicates), prefixed with a symbol. If a predicate prefixed with a \texttt{++} is added to the belief store (either as a fact or rule), then the rule might stop firing. If a predicate prefixed with \texttt{-{}-} is removed from the belief store, then the rule might stop firing. For example, consider the below procedure:
\begin{verbatim}
proc1(){
a & b ~> m1
d & e ~> m2
f & g ~> m3
true ~> m4
}
\end{verbatim}
The ``local dependent predicates'' for the first rule are \texttt{[-{}-a, -{}-b]}.\\
For the second rule they are \texttt{[-{}-d, -{}-e, ++a, ++b]}.\\
For the third rule they are \texttt{[-{}-f, -{}-g, ++a, ++b, ++d, ++e]}.\\
For the last rule they are \texttt{[++a, ++b, ++d, ++e, ++f, ++g]}.\\

To check if a rule must continue firing, the union of all of the local dependent predicates of all of the rule's parent calls is calculated to produce the ``dependent predicates'' of the current state of the program\citep{Clark2014}.

To demonstrate this, consider the following program:
\begin{verbatim}
proc1(){
a & b ~> c
d & e ~> f
f ~> proc2()
true ~> g
}

proc2(){
k ~> l
m ~> n
true ~> q
}
\end{verbatim}
Say the program was called by calling the \texttt{proc1} procedure. If the second rule of \texttt{proc2} is currently firing (as a result of the third rule of \texttt{proc1} being fired), the ``dependent predicates'' can be determined by finding the union of the local dependent predicates for the third rule of \texttt{proc1} and the second rule of \texttt{proc2}.\\
These are:\\
\texttt{proc1}, rule 3: \texttt{[++a, ++b, ++d, ++e, -{}-f]}.\\
\texttt{proc2}, rule 2: \texttt{[++k, -{}-m]}.\\

So the dependent predicates are \texttt{[++a, ++b, ++d, ++e, -{}-f]} \(\cup\) \texttt{[++k, -{}-m]} \(=\) \texttt{[++a, ++b, ++d, ++e, -{}-f, ++k, -{}-m]}.\\

\subsection{Evaluation of \textsc{TR} conditions}
In Nilsson's 2001 paper\citep{Nilsson2001} introducing the ``Triple Tower Architecture'', the inference rules in the Perception Tower were used to populate the Model Tower with derived knowledge but the way in which this is done is not elaborated on. However, there are papers have given formal semantics for \textsc{TR}, such as \citep{dongol2014reasoning} in a temporal logic and \citep{kowalski2012teleo} as abductive logic programming. Nilsson mentions that the ``not'' expression in \textsc{TR} means ``negation-as-failure'' so (given the fact that the rules in the ``Perception Tower'' resemble Horn clauses) it is reasonable to assume that \textsc{TR} uses SLDNF (Selective Linear Definite clause resolution with Negation by Failure)\citep{sergot2010nbf}.

\section{\textsc{TeleoR}/Qulog}
\textsc{TeleoR} is a language devised by Keith Clark and Peter Robinson as an extension of the original language \textsc{TR}\citep{Clarka}. The declarative logic/functional language used to express guard conditions, relations (predicates) and functions is called Qulog. The added features are as follows:
\begin{itemize}
\item procedures and the BeliefStore language (Qulog) are typed and allow for higher order programming;
\item timed action sequences that can specify a sequence of actions to be executed cyclically;
\item while/until rules which allow the programmer to provide additional conditions under which the rule can fire;
\item wait/repeat actions can be given, which lets a discrete action be repeated if it has not resulted in the firing of another rule within some period of time. This is useful in cases where an action does not have the desired effect first time (e.g. if something in the robot mechanism jams);
\item actions are provided that dynamically modify the BeliefStore, so that that agent can ``remember'' and ``forget'' facts;
\item the ability to link BeliefStore updates and message send actions to any other agents with any rule action, allowing for inter-agent communication.
\end{itemize}

\subsection{Two-tower architecture}
The high-level design of \textsc{TeleoR} is different to that of \textsc{TR} in that it does not include a truth maintenance system. The three-tower architecture becomes a \textbf{two-tower architecture}, with two towers:
\begin{itemize}
\item the \textbf{BeliefStore tower} - this deduces truths about the world from the incoming percepts, essentially having the same functions as the perception tower in \textsc{TR}. 
\item the \textbf{Action tower} - this performs the same role as the action tower in \textsc{TR}, although the language used to determine the actions to perform is different, with some added features.
\end{itemize}

\begin{figure}[ht!]
\centering
\includegraphics[width=90mm]{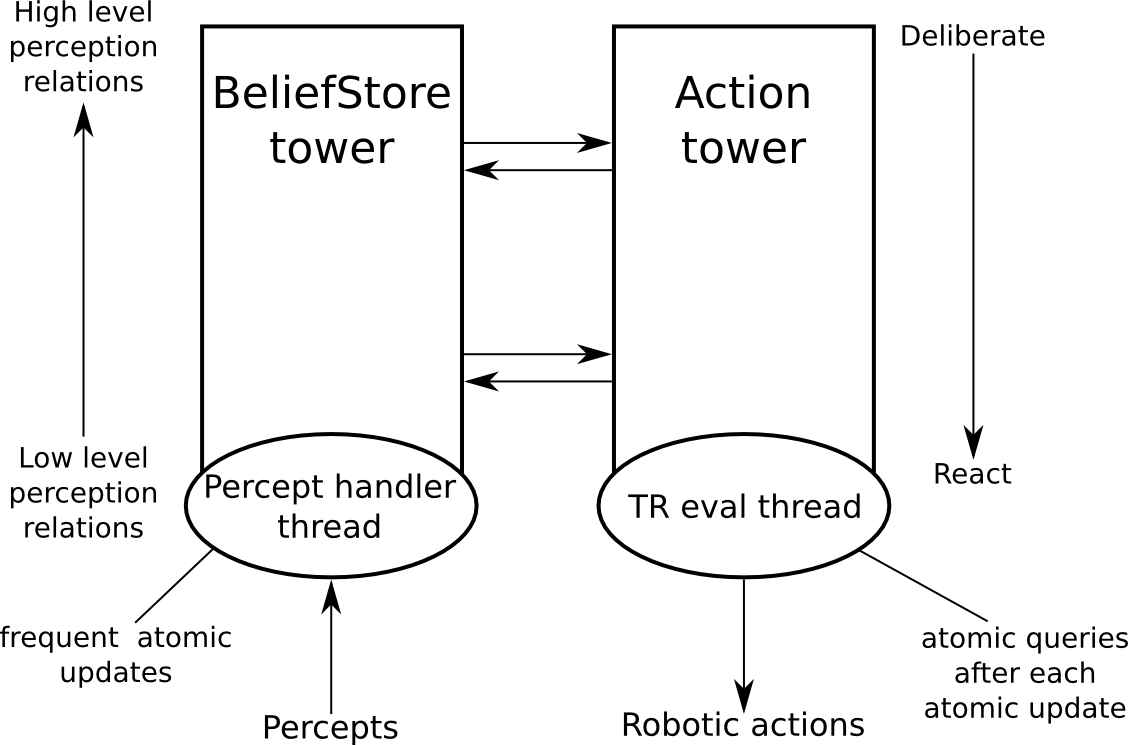}
\caption{Diagram of the two-tower architecture \label{twotowerdiagram}}
\end{figure}

\subsection{\textsc{TeleoR} Syntax}
Now for the syntax of the \textsc{TeleoR} language. A procedure takes the following form. Like in the description of \textsc{TR}, \texttt{procedure\_name} is the name of the procedure and \texttt{param\_1, ... param\_k} are the names of the parameters. Added to the procedure definition is a type declaration, where \texttt{t\_i} is the type of parameter \texttt{param\_i}. \\

\begin{verbatim}
procedure_name : (t_1, ..., t_k) ~>
procedure_name(param_1, ... , param_k){
  G_1 while WC_1 min WT_1 until UC_1 min UT_1 ~> R_1
  G_2 while WC_2 min WT_2 until UC_2 min UT_2 ~> R_2
  ...
  G_n while WC_n min WT_n until UC_n min UT_n ~> R_n
}
\end{verbatim}

The first line is new to \textsc{TeleoR}, it is a type signature. It states which types of parameters can be passed to the procedure. The way in which types are checked and how this aids development will be discussed in another section. The left-hand side of each rule has also become more complex, these are the aforementioned ``while/until'' conditions. These are explained in the ``Guard conditions'' section. If a part of the guard contains vacuous constraints (i.e. constraints that do not alter the behaviour of the rule) then they can be omitted.  The right-hand sides of the rules (\(R_1\) to \(R_n\) are also more complex than for \textsc{TR}. Not only can they contain a tuple of action primitives or a call to a procedure, they can contain expressions of the form:
\begin{verbatim}
A_1 for N_1; A_2 for N_2; ... ;A_m for N_m
A wait N_1 repeat N_2
\end{verbatim}
where \(A_i\) are either primitive actions to be executed in parallel (separated with commas) or single procedure calls. The semantics of the timed sequence actions and the wait/repeat actions will be explained later. 

\subsection{The BeliefStore in \textsc{TeleoR}}
``BeliefStore'' is a term used to describe whatever the source of percepts and beliefs in the teleo-reactive system is. In the case of \textsc{TeleoR} it is the language QuLog, which was also developed by Clark and Robinson. The language itself will be described in more detail later in this chapter\citep{Clark2014} in Section \ref{subsec:relationdefinitions}, \ref{subsec:functiondefinitions} and \ref{subsec:typesystem}.

\subsection{Extensions to \textsc{TR} rules}
\subsubsection{Guard conditions}
In \textsc{TeleoR}, the language of guard conditions has been made more expressive. This has been done by the introduction of \textbf{while} and \textbf{until} conditions. These are of the form\citep{Clarka}:
\begin{verbatim}
G while WC min WT until UC min UT ~> A
\end{verbatim}
This means that once the rule has begun to fire, it will continue to fire while (Equation \ref{continuationcondition1}) holds (the \textit{continuation condition}).
\begin{equation}
\mathtt{inferable}(G) \lor ( (\mathtt{inferable}(WC) \lor \neg \mathtt{expired}(WT)) \land (\neg \mathtt{inferable}(UC) \lor \neg \mathtt{expired}(UT)) )
\label{continuationcondition1}
\end{equation}
where $\mathtt{inferable}(X)$ means that the query $X$ is currently inferable and $\mathtt{expired}(T)$ means that more than $T$ seconds have elapsed since the rule started firing. If \texttt{WC} or \texttt{UC} are not given, they default to \texttt{false}. If \texttt{WT} or \texttt{UT} are not given, they default to \texttt{0}. If \texttt{X} is 0, then $\mathtt{expired}(X)$ wil always be \texttt{true}. So, based on the original form, the following variations are possible:\\

\begin{verbatim}
G while WC min WT until UC
\end{verbatim}
has the corresponding (simplified) continuation condition (Equation \ref{continuationcondition2}).
\begin{equation}
\mathtt{inferable}(G) \lor ( (\mathtt{inferable}(WC) \lor \neg \mathtt{expired}(WT)) \land \neg \mathtt{inferable}(UC) )
\label{continuationcondition2}
\end{equation}

\begin{verbatim}
G while WC until UC min UT
\end{verbatim}
has the corresponding (simplified) continuation condition (Equation \ref{continuationcondition3}).
\begin{equation}
\mathtt{inferable}(G) \lor ( \mathtt{inferable}(WC) \land (\neg \mathtt{inferable}(UC) \lor \neg \mathtt{expired}(UT)) )
\label{continuationcondition3}
\end{equation}

\begin{verbatim}
G while WC min WT
\end{verbatim}
has the corresponding (simplified) continuation condition (Equation \ref{continuationcondition4}).
\begin{equation}
\mathtt{inferable}(G) \lor (\mathtt{inferable}(WC) \lor \neg \mathtt{expired}(WT))
\label{continuationcondition4}
\end{equation}

\begin{verbatim}
G while WC until UC
\end{verbatim}
has the corresponding (simplified) continuation condition (Equation \ref{continuationcondition5}).
\begin{equation}
\mathtt{inferable}(G) \lor ( \mathtt{inferable}(WC)  \land \neg \mathtt{inferable}(UC) )
\label{continuationcondition5}
\end{equation}

\begin{verbatim}
G
\end{verbatim}
has the corresponding (simplified) continuation condition (Equation \ref{continuationcondition6}).
\begin{equation}
\mathtt{inferable}(G)
\label{continuationcondition6}
\end{equation}

The effect of \textbf{while} and \textbf{until} conditions is confined to the procedure-level. This means that if a rule in some procedure is fired but the procedure stops being called, the rule's action will always stop firing, regardless of any while/until conditions that the rule may have\citep{qulogmanual}.\\

\begin{figure}[h!]
\centering
\includegraphics[scale=0.5, clip, trim=0cm 15cm 0cm 4cm]{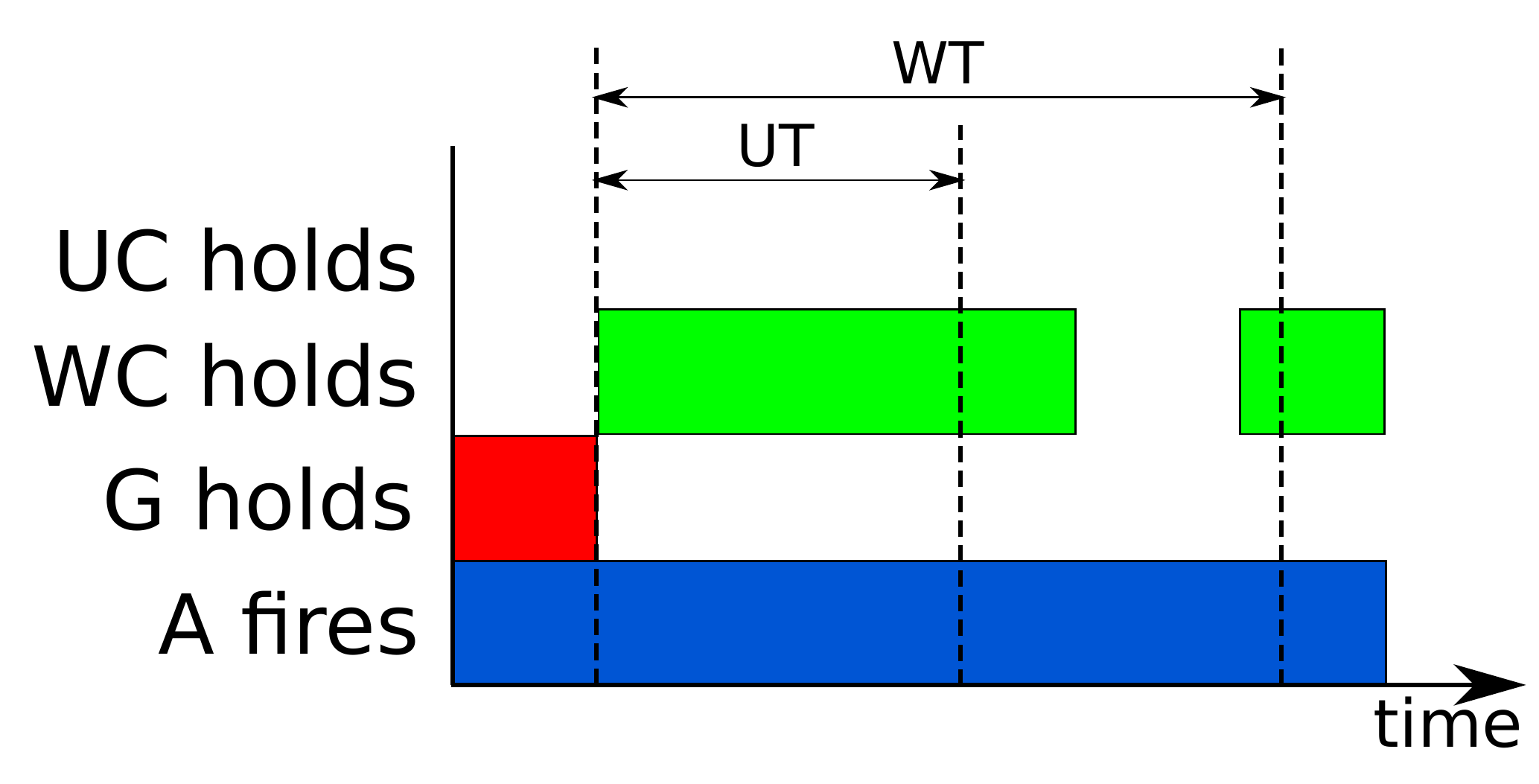}
\caption{Example rule firing}
\label{fig:whileuntil1}
\end{figure}

\begin{figure}[h!]
\centering
\includegraphics[scale=0.5, clip, trim=0cm 15cm 0cm 4cm]{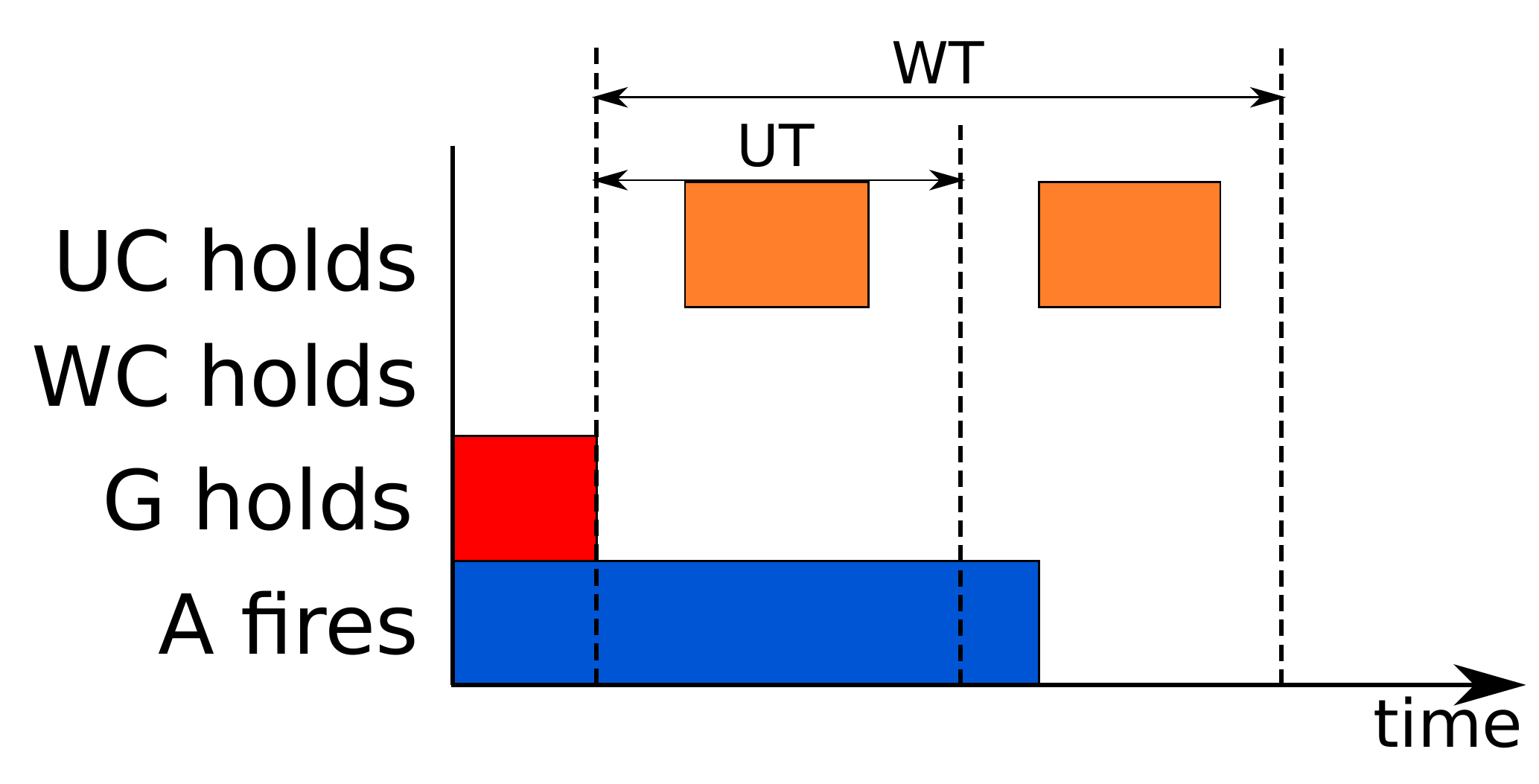}
\caption{Example rule firing}
\label{fig:whileuntil2}
\end{figure}

The behaviour of the while/until rules can be illustrated with diagrams showing example rule firings. Consider the rule \verb|G while WC min WT until UC min UT ~> A|, where \texttt{WT > UT}. In the first case (Figure \ref{fig:whileuntil1}), the rule fires, then \texttt{WC} becomes inferable just after \texttt{WT} seconds, then the rule stops firing after \texttt{WC} stops being inferable. In the second case (Figure \ref{fig:whileuntil2}), the rule fires then \texttt{UT} becomes inferable, but it does not stop the rule from firing, then it stops being inferable. Then it becomes inferable again, but because this happens after \texttt{UT} seconds, it causes the rule to stop firing.\\

\subsubsection{Timed sequences of actions}
\begin{verbatim}
A_1 for T_1; T_2 for T_2; ... ;A_m for T_m
\end{verbatim}
The above action performs m actions in sequence. A\_1 is performed for N\_1 seconds and then A\_2 will be performed for N\_2 seconds and so on until A\_m then the agent will start at A\_1 again \citep{Clark}. A\_i can be a tuple of primitive actions or a procedure call. If \texttt{for T\_m} is missing, then the corresponding action will run indefinitely, until the rule stops firing.\\
For example, given a robot with a \texttt{turn/1} durative action that takes a direction as input (\texttt{direction ::= left | right}) and a \texttt{move\_forward/0} durative action, a task could be defined that causes the robot to move in a zig-zag motion:\\
\begin{verbatim}
zigzag : () ~>
zigzag(){
true ~> move_forward, turn(left) for 0.2;
        move_forward, turn(right) for 0.2
}
\end{verbatim}

In this example, the robot moves forward and turns left for 0.2 seconds, then moves forward and turns right for 0.2 seconds, then moves forward and turns left for 0.2 seconds and so on.\\

Timed sequence actions can also be used to tell the agent to do some things, then do one thing forever. For example, the task \texttt{move\_forward\_then\_turn\_left} could be defined as follows:\\
\begin{verbatim}
move_forward_then_turn_left : () ~>
move_forward_then_turn_left(){
true ~> move_forward for 1;
        turn(left)
}
\end{verbatim}
This task tells the agent to move forward for 1 second, then turn left indefinitely.\\

\subsubsection{wait ... repeat actions}
The wait ... repeat construct allows an action to be tried, then retried a number of times at a given interval. This can be useful if the agent has to perform an action that may not succeed and may have to be retried. For example, if a door-opening grabs and turns a handle on a door, then it could be programmed to retry if the door does not open. It has the following syntax:
\begin{verbatim}
A wait T repeat R
\end{verbatim}
The above code will perform the action \texttt{A} and after \texttt{T} seconds the action \texttt{A} will be tried again a maximum of \texttt{R} times unless another rule is chosen. If no rule is chosen after \texttt{R} repeats, an error is generated \texttt{action\_failure} which appears as a percept which can be queried by the parent procedure \citep{Clarka, qulogmanual}.\\

For example, if an agent is programmed to open a door by pushing it (where \texttt{push} is a discrete action), it may not succeed on the first try (e.g. the door is jammed or is locked). So the wait/repeat construct can be used to retry the action a given number of times. In this example, if the \texttt{do\_things} procedure is called then if agent sees a closed door then it will call \texttt{open\_the\_door} which (unless an open door is already observed) will cause the robot to push on the door, wait one second, then push on the door three more times at one second intervals. If no other rule is fired by that point, a special \texttt{action\_failure} percept will be remembered, which will cause the first rule in \texttt{do\_things} to fire. The robot can now deal with the fact that \texttt{open\_the\_door} had failed, e.g. by using a key (to unlock the door).\\

\begin{verbatim}
do_things(){
action_failure ~> use_key()
see(closed_door) ~> open_the_door()
true ~> ()
}

open_the_door(){
see(open_door) ~> ()
true ~> push wait 1 repeat 3
}
\end{verbatim}

\subsubsection{remember, forget actions}
The \textsc{TeleoR} language provides primitive actions to add and remove beliefs dynamically from the BeliefStore, which can be used to give the agent some `memory' or state. This is done with the \texttt{remember} and \texttt{forget} actions. The remember action adds a belief to the BeliefStore and the forget action removes a belief. These beliefs can be queried in the guards of the rules like the percepts\citep{Clark2014}. The belief must have been declared as a \texttt{belief} (with a type) \citep{qulogmanual}.\\

\subsection{Example \textsc{TeleoR} Programs}
I will now give some example programs, that illustrate how \textsc{TeleoR} programs can be designed.

Teleo-reactive programming is ideal for expressing the behaviour of systems whose behaviour reacts continuously to the outside world. One such system is a thermostat. This is a machine that switches a heater on and off, in order to keep a room or other space at a certain temperature. In its simplest form, it consists of a thermometer that turns on a switch if the temperature is above (or below) a certain desired level. If the temperature is too low, the heating is turned on, if it is too high then the heating is turned off. This behaviour has been expressed in the following teleo-reactive program.

\begin{verbatim}
discrete turn_on_heating : (),
         turn_off_heating : ()

percept  is_too_cold : ()

thermostat_task : () ~>
thermostat_task(){
is_too_cold ~> turn_on_heating
true ~> turn_off_heating
}
\end{verbatim}
This program has two actions \texttt{turn\_on\_heating} and \texttt{turn\_off\_heating}, and one percept \texttt{is\_too\_cold} which becomes inferable if the temperature of the room is too cold. In order to run the program, the \texttt{thermostat\_task} procedure is called, with no arguments.

The next program generalises the above program, to take the desired temperature as an argument. 

\begin{verbatim}
discrete turn_on_heating : (),
         turn_off_heating : ()

percept temperature : (num)


regulate_temperature : (num) ~>
regulate_temperature(Target){
temperature(Temperature) & Temperature < Target ~> turn_on_heating
true ~> turn_off_heating
}
\end{verbatim}

The \texttt{is\_too\_cold} percept from the first example has been replaced with a \texttt{temperature} percept which, as stated in its type signature, has one term which is of type \texttt{num} (a number). If the percept \texttt{temperature(T)} is inferable, it means that the temperature is \texttt{T} degrees Celsius.
The \texttt{regulate\_\-temperature} task also takes a parameter, which is the temperature that the thermostat must maintain. So the temperature that the thermostat can be chosen by whatever calls the procedure.
The first rule in the procedure states that if a certain temperature is sensed, if that temperature is lower than the desired temperature, then turn on the heating. The second rule covers all other cases (i.e. no temperature detected or the temperature was equal or greater than the desired temperature), in which case the heating is turned off. Now we have a general-purpose procedure for a thermostat that can maintain any temperature. This can then be re-used in another part of the program, as the following example will show:

\begin{verbatim}
discrete turn_on_heating : (),
         turn_off_heating : ()

percept temperature : (num),
        person_in_room : ()


thermostat_behaviour : () ~>
thermostat_behaviour(){
person_in_room ~> regulate_temperature(28)
true ~> regulate_temperature(18)
}

regulate_temperature : (num) ~>
regulate_temperature(Target){
temperature(Temperature) & Temperature < Target ~> turn_on_heating
true ~> turn_off_heating
}
\end{verbatim}

This program introduces a new procedure \texttt{thermostat\_behaviour} and a new percept \texttt{person\_\-in\_\-room} which is inferable if a person is detected in the room. The \texttt{thermostat\_behaviour} procedure states that if a person is sensed in the room, maintain the temperature at 28 degrees, otherwise maintain it at 18 degrees. It shows how (like in other programming paradigms) procedures can be re-used to incrementally build up more complex behaviour. 
Another thing to note is that because of the type signatures given by the programmer, this program can be proven at compile time to never produce a run-time type error (unless provided with invalid percepts or arguments). The signature of \texttt{regulate\_temperature} states that the argument must be a number, in both cases in \texttt{thermostat\_behaviour} it is called with numbers (18 and 25). Given that, it is guaranteed that Target will not cause a type error when compared with Temperature (because Temperature is of type num, given the type signature of \texttt{temperature}).

\subsection{Relation definitions} \label{subsec:relationdefinitions}
Relations can be defined in QuLog using two constructs: Unguarded Rules and Guarded Rules. The definition of the syntax of these rules is copied from the paper describing QuLog \citep{Clark}.\\
\textbf{Unguarded Rules} have the following syntax:
\begin{verbatim}
Head
or
Head <= ComplexConj
\end{verbatim}
where \texttt{Head} is a head predication of the form \texttt{rel(Arg1, \ldots , Argk)} where k > 0.\\
A \texttt{ComplexConj} has the form \texttt{Cond1 \& ...  Condn}, \(n \geq n\), with \texttt{Condi} being:
\begin{itemize}
\item a body predication \texttt{RelExp(Exp1, ... , Expk)}, \(k \geq 0\), where each \texttt{Expi} is an expression - a term that may contain function calls. \texttt{RelExp} is an expression returning a k-ary relation \texttt{rel'} such that the values of the argument expressions will satsify the mode and type constraints of \texttt{rel'} when it is called.
\item a body predication prefixed with \texttt{not}, the QuLog negation-as-failure
\item a body predication prefixed with \texttt{once}, indicating that only one successful evaluation should be found
\item an expression value unification \texttt{Exp1 = Exp2}
\item a non-deterministic pattern match \texttt{Exp =? PtnTerm}
\item a meta-call \texttt{call Call}, of type \texttt{!relcall}
\item a universally quantified implication (a forall) of the form: \texttt{forall V1, ... , Vj} (exists $EVarsSeq_1$ $SimpleConj_1$ \texttt{=>} $EVarsSeq_2$ $SimpleConj_2$), $j \geq 1$\\
The \texttt{Vi} variables are universally quantified over the implication. The sequence of variables of \texttt{EVarsSeqi} are existentially quantified over \texttt{SimpleConj1} and \texttt{SimpleConj2}.
\end{itemize}
A \texttt{SimpleConj} is a \texttt{ComplexConj} that does not contain any \texttt{forall}.\\
\textbf{Guarded Rules} are of the following form:
\begin{verbatim}
Head :: Commit <= Body
\end{verbatim}
\texttt{Commit} is a \texttt{SimpleConj} and \texttt{Body} is a \texttt{ComplexConj}.\\

The Commit is a test that, if passed, no other definitions will of the relation will be considered. So the above definition is similar to \texttt{Head :- Commit, !, Body} in Prolog, but Qulog does not have the cut (\texttt{!}) operator.\\

Every relation has a \textbf{Relation Type Declaration}, which states the type and mode of the terms of the relation. It is of the form:
\begin{verbatim}
rel: (m1t1, ... , mktk) <=
\end{verbatim}
where \texttt{rel} is the name of the relation, \texttt{ti} is the ith type expression and each \texttt{mi} is one of the three prefix mode annotations !, ?, ??. If the type is omitted, then the type \texttt{term} is used (i.e. any type of term is accepted). The postfix \texttt{?} is equivalent to \texttt{??} and the prefix \texttt{!} can be dropped\citep{Clark}.

\subsection{Function definitions}\label{subsec:functiondefinitions}
Functions can be expressed in two ways:
\begin{verbatim}
fun(Arg0, ... , Argk) -> Exp
or
fun(Arg0, ... , Argk) :: SimpleConj -> Exp
\end{verbatim}
Like in the rule definitions, the \texttt{SimpleConj} is the ``commit test''. \texttt{fun(Arg0, ... , Argk)} is the head of the function definition. To evaluate a function call, the expressions \texttt{Arg0, ... , Argk} are first evaluated. Then, the first rule for \texttt{fun} with a head that matches the argument call and passes the commit test (if any) determines the call's value. This is the value of the right hand side expression \texttt{Exp}, made ground by the argument call and any values from the evaluation of the commit test\citep{Clark}.

\subsection{Type system}\label{subsec:typesystem}
Qulog allows the programmer to specify the types of variables, percepts, beliefs, actions and procedures in order to catch potential errors at compile time. In addition, relations/predicates can be assigned moded types which state whether the input/output of each term must be ground.\\

To demonstrate type checking, consider the following program. It has two declared types ``thing'' and ``direction''. The percept ``see'' takes three terms of type ``thing'', ``direction'' and ``num''. When this program is loaded by Qulog, it will trigger a type error because the atom ``dog'' is not declared as belonging to the type ``thing''. 
\begin{verbatim}
thing ::= box | shoe | cat
direction ::= left | centre | right

durative move_forward : ()
         turn_right : ()

percept see : (thing, direction, num)

proc(){
see(box, left, 10) ~> move_forward
see(dog, right, 2) ~> turn_right, move_forward
}
\end{verbatim}

Qulog also performs run-time type checking, for example if a percept was declared as having a certain type in the source program then if a percept with invalid type is received by the agent, an error will be fired.

\subsubsection{Defining types}
In QuLog, the type system was used to check if the terms of the predicates and actions were valid. QuLog was ``statically typed'', which means that type checking took place at compile time. The types of percepts, beliefs, actions and procedures had to be given manually by the programmer. The language only performed type checking, not type inference.\\

The language consists of built-in types, from which the user can define new types. These are:
\begin{itemize}
\item \(string\) - a string, enclosed in double quotes
\item \(integer\) - an integer, i.e. a whole number
\item \(natural\) - a natural number, i.e. an integer \(N\) where \(N \geq 0\).
\item \(number\) - a real (integer, floating point) number
\item \(atom\) - an atom
\item \(type\) - a type
\item \(top\) - the type at the top of the type hierarchy
\item \(bottom\) - the type at the bottom of the type hierarchy
\end{itemize}

\begin{figure}[H]
\centering
\begin{tikzpicture}[level/.style={sibling distance=20mm/#1}]
\node (atomic){$atomic$}
  child {node (num) {$num$}
    child {node (int) {$int$}
      child {node (nat) {$nat$}
      }
    }
  }
  child {node (atom) {$atom$}}
  child {node (string) {$string$}}
;
\path (atomic) -- (atom);
\path (atomic) -- (string);
\path (atomic) -- (num);
\path (num) -- (int);
\path (int) -- (nat);
\end{tikzpicture}
\caption{The type hierarchy of the built-in QuLog types}
\label{fig:typehierarchy}
\end{figure}
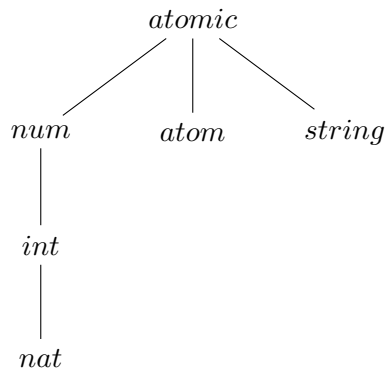

\begin{figure}[H]
\centering
\begin{tikzpicture}[level/.style={sibling distance=20mm/#1}]
\node (atomic){$atomic$}
  child {node (num) {$num$}
    child {node (int) {$int$}
      child {node (nat) {$nat$}
      }
    }
  }
  child {node (atom) {$atom$}
    child {node (legume) {$legume$}}
  }
  child {node (string) {$string$}}
;
\path (atomic) -- (atom);
\path (atomic) -- (string);
\path (atomic) -- (num);
\path (num) -- (int);
\path (int) -- (nat);
\path (atom) -- (legume);
\end{tikzpicture}
\caption{The type hierarchy, after adding a new user-defined type \(legume\).}
\label{fig:typehierarchypea1}
\end{figure}
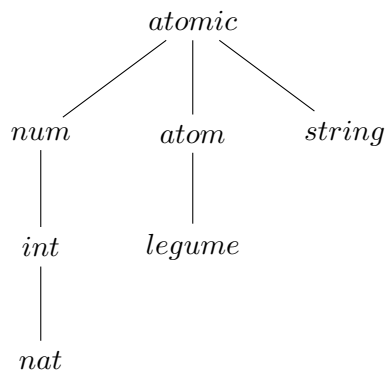

The types are arranged in a partial order: if something belongs to a type, then it also belongs to all of its parent types. This relation is written with a \(\geq\) or \(>\) symbol. It has the following properties:
\begin{itemize}
\item transitivity - $A \geq B \land B \geq C \Longrightarrow A \geq C$
\item equality - \(A \geq A\)
\item bottom - \(T > Bottom\), the bottom type is below all types
\item top - \(Top > T\), the top type is above all types
\end{itemize}
The ordering of types for the built-in types is:
\begin{itemize}
\item \(atomic > num > int > nat\)
\item \(atomic > atom\)
\item \(atomic > string\)
\end{itemize}
In this case, \(atomic\) is the parent of all types, \(num\) is the parent of \(int\) and so on. This can be depicted as a tree, in Figure \ref{fig:typehierarchy}.
The user can define a new type in three different ways: as a disjunction of atoms, a disjunction of types or a range type. A disjunction of atoms definition states that each atom in the definition is a member of the new type. For example:
\begin{verbatim}
legume ::= haricotverts | cannellini | azuki | pea
\end{verbatim}
The left-hand side of the ``::='' operator is the new type and the constituent atoms are separated by a ``|''. The above code defines a new type \(legume\) which has four members, \(haricotverts\), \(cannellini\), \(azuki\) and \(pea\). As this type is a disjunction of atoms, the type \(legume\) is below \(atom\) in the type hierarchy. So after evaluating the above line, the type hierarchy now looks like the one in Figure \ref{fig:typehierarchypea1}.

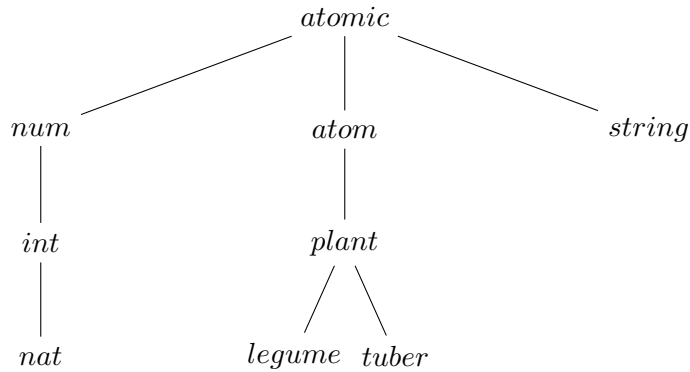
\begin{figure}[H]
\centering
\begin{tikzpicture}[level/.style={sibling distance=40mm/#1}]
\node (atomic){$atomic$}
  child {node (num) {$num$}
    child {node (int) {$int$}
      child {node (nat) {$nat$}
      }
    }
  }
  child {node (atom) {$atom$}
    child {node (plant) {$plant$}
      child {node (legume) {$legume$}}
      child {node (tuber) {$tuber$}}
    }
  }
  child {node (string) {$string$}}
;
\path (atomic) -- (atom);
\path (atomic) -- (string);
\path (atomic) -- (num);
\path (num) -- (int);
\path (int) -- (nat);
\path (atom) -- (plant);
\path (atom) -- (legume);
\path (atom) -- (tuber);
\end{tikzpicture}
\caption{The type hierarchy, after adding two types \(legume\) and \(tuber\) and a disjunction of types \(plant\).}
\label{fig:typehierarchypea2}
\end{figure}

The second way that new types can be defined is as a disjunction of types. The new type can be considered the ``parent'' type, which has many ``child'' types. The parent type is above every child type in the type hierarchy and below \(atom\). For example, imagine that the \(legume\) type from above is in the type hierarchy and a \(tuber\) type has also been defined. A disjunction of types could be defined as follows:
\begin{verbatim}
plant ::= legume || tuber
\end{verbatim}
This says that the new type \(plant\) contains all items in \(legume\) and \(tuber\). The left hand side of the ``::='' operator is the new parent type and the child types (\(N \geq 2\)) are separated by ``||''. In other words, if an atom is in \(legume\) or \(tuber\), then it is also in \(plant\). So the type hierarchy looks like the one in Figure \ref{fig:typehierarchypea2}.

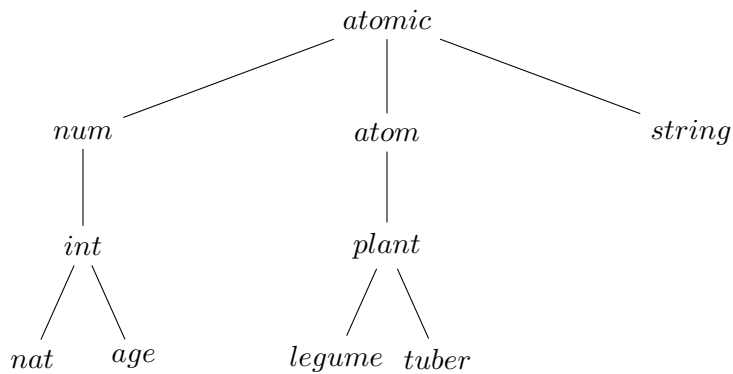
\begin{figure}[H]
\centering
\begin{tikzpicture}[level/.style={sibling distance=40mm/#1}]
\node (atomic){$atomic$}
  child {node (num) {$num$}
    child {node (int) {$int$}
      child {node (nat) {$nat$}}
      child {node (age) {$age$}}
    }
  }
  child {node (atom) {$atom$}
    child {node (plant) {$plant$}
      child {node (legume) {$legume$}}
      child {node (tuber) {$tuber$}}
    }
  }
  child {node (string) {$string$}}
;
\path (atomic) -- (atom);
\path (atomic) -- (string);
\path (atomic) -- (num);
\path (num) -- (int);
\path (int) -- (nat);
\path (int) -- (age);
\path (atom) -- (plant);
\path (atom) -- (legume);
\path (atom) -- (tuber);
\end{tikzpicture}
\caption{The type hierarchy, after adding two types \(legume\) and \(tuber\), a disjunction of types \(plant\) and a range type \(age\).}
\label{fig:typehierarchypea3}
\end{figure}

The third way of defining a type in QuLog is a range of integers. This define a new type that is a child of the built-in \(int\) type. A maximum and minimum integer is given. For example, if one wants to define a new type \(age\) with a lower bound \(0\) and an upper bound \(120\) then the definition would look like the following:
\begin{verbatim}
age ::= (0 .. 120)
\end{verbatim}
This adds a new type to the type hierarchy, so that it looks like Figure \ref{fig:typehierarchypea3}

\subsubsection{Type checking}
The paper by Clark \& Robinson \citep{Clark} gives rules to infer, check and simplify types in QuLog. It also gives an informal description of how to type check a QuLog program, using these rules.

\subsubsection{Moded types}
The way that predicates in languages like Prolog and QuLog are evaluated means that they can be defined in a way that allows them to be called in ``either direction''. For example, consider the predicate ``append'' (from the SWI Prolog distribution) that has three terms, ``List1'', ``List2'' and ``List1AndList2''. The documentation states that ``List1AndList2 is the concatenation of List1 and List2''. Depending on which terms are given as ground, this can be called in multiple different ways.

\begin{verbatim}
?- append([1,2], [3,4], Z).
Z = [1, 2, 3, 4].
\end{verbatim}
The above query finds the list that is the concatenation of \([1,2]\) and \([3,4]\).

\begin{verbatim}
?- append([1,2], Y, [1,2,3,4]).
Y = [3, 4].
\end{verbatim}
The above query finds the list that, when concatenated to the end of \([1,2]\), produces \([1,2,3,4]\).

\begin{verbatim}
?- append(X, [3,4], [1,2,3,4]).
X = [1, 2] ;
false.
\end{verbatim}
The above query finds the list that, when \([3,4]\) is concatenated to the end of it, produces \([1,2,3,4]\).

\begin{verbatim}
?- append(X, Y, [1,2,3,4]).
X = [],
Y = [1, 2, 3, 4] ;
X = [1],
Y = [2, 3, 4] ;
X = [1, 2],
Y = [3, 4] ;
X = [1, 2, 3],
Y = [4] ;
X = [1, 2, 3, 4],
Y = [] ;
false.
\end{verbatim}
The above query finds all of the possible lists that concatenate to produce \([1,2,3,4]\).

The above examples demonstrate that every argument/term of the ``append/3'' predicate can be given as a ground value (input) such as a list, string, number or atom or a variable (output). The unification algorithm used by Prolog determines the possible values of all of the remaining variables. However, some predicates require that certain arguments be given. Whether or not an argument must be given is its ``mode''. In Prolog, the mode is given in the documentation for the predicates for the programmer's information. QuLog improves on this by checking the modes of predicates at compile time as part of the type-checking process. The mode of the predicates is stated by putting one of three symbols before the declared type name:
\begin{itemize}
\item !, ground input - the term must be ground before the predicate is called;
\item ?, ground output - the term will be ground after the predicate is called;
\item ??, unconstrained output - the term does not have to be ground after the predicate is called.
\end{itemize}
Then, type checking rules are used to confirm that the modes are given correctly. The rules are given as follows:
\begin{equation}
\infer{!T_1 \leq_{\textbf{m}} !T_2}{T_2 \leq T_1}
\end{equation}
\begin{equation}
\infer{?T \leq_{\textbf{m}} ??T}{}
\end{equation}
\begin{equation}
\infer{?T_1 \leq_{\textbf{m}} !T_2}{T_2 \leq T_1}
\end{equation}
\begin{equation}
\infer{??T_1 \leq_{\textbf{m}} !T_2}{T_2 \leq T_1}
\end{equation}

These definitions introduce a new comparison operator (lte with m). \(T \leq_{\textbf{m}} T'\) means that \(T\) is less than or equal to \(T'\), as a moded type. 

\subsection{Evaluation of conditions in Qulog}
In order to determine which rule to fire, the conditions on the left-hand side of the rules have to be evaluated. The conditions make up a QuLog query. If the query contains variables, then the instantiations of the variables which cause the query to be true will also have to be found. This method is used to evaluate the guard conditions, while conditions and until conditions of the left-hand side of the \textsc{TR} procedure rules, as well as individual queries made in the QuLog terminal. A guard condition is a \texttt{ComplexConj} object, as defined in Section \ref{subsec:relationdefinitions}\citep{qulogmanual, Clark}.

\subsubsection{Unification}
The unification problem is ``given two terms containing some variables, find the simplest substitution (an assignment of variables to terms) that makes the terms equal''. Several algorithms exist for unification, the first one by Robinson in 1965 \citep{Robinson1965} is simple but inefficient, since then faster algorithms have been devised such as those by Martelli \& Montanari in 1976 \citep{martelli1976unification} and 1982 \citep{Martelli1982} and Paterson \& Wegman in 1976 \citep{paterson1976linear}.

\subsection{Concurrency}
Two forms of concurrency are provided by Qulog - concurrency within the agent and concurrency between agents. The first allows the agent to work towards fulfilling multiple goals (perform tasks) concurrently, the second lets agents communicate (and thereby collaborate) with each other\citep{qulogmanual}.

\subsubsection{Concurrency within the agent}
Concurrency within the agent is necessary because an agent may need to perform several tasks at the same time. Each task may have control of some resources that are shared among the tasks. For example, an agent could control a robot arm and it could be tasked with building several towers - but the robot arm can only build one tower at a time. So the robot arm is a resource, the control of which is shared among the tasks\citep{qulogmanual}. The details of how this is achieved is beyond the scope of this report.

\subsubsection{Concurrency between agents}
To start an agent, use the action \texttt{start\_agent(Name, Handle, Convention)}, where \texttt{Name} is the name (atom) of the new agent and \texttt{Handle} is the message address of the interface or simulation with which the robot will interact. \texttt{Convention} is the percept update convention being used. This can be one of three values:
\begin{itemize}
\item \texttt{all} - the interface/simulation sends all percepts when there is an update;
\item \texttt{updates} - the interface/simulation sends only the percepts that have changed when there is an update;
\item \texttt{user} - the interface/simulation sends percepts in an application-specific way, in which case the action \texttt{handle\_percepts\_} will need to be defined.
\end{itemize}
To kill (terminate) an agent, use the action \texttt{kill\_agent(Name)}, where \texttt{Name} is the name (atom) of the agent to be killed. This name is the same as the one used when starting the agent with \texttt{start\_agent}\citep{qulogmanual}.

\newpage
\section{Pedro}
The decision-making part of the robot (Qulog/\textsc{TeleoR}) needs to communicate with other programs. It will need to communicate with whatever the source of percepts is (simulation, sensors) and communicate with whatever the recipient of the actions is (simulation, actuators). It may also need to communicate with other robots and possibly receive information from their sensors or commands. Pedro provides a protocol that can be used to send and receive all of this information\citep{Robinson2009, qulogmanual}.

\subsection{How agents connect to Pedro}
The process for connecting to Pedro is as follows\citep{Robinson2009}:
\begin{enumerate}
\item Create a socket in the client.
\item Use \texttt{connect} to connect the socket to the Pedro server. The default port is 4550.
\item Read an IP address and two ports (sent by the server as a newline terminated string). The IP address is to be used for the connection to the server. The two ports are for connecting two sockets (for acknowledgements and for data).
\item Close the socket.
\item Create a socket in the client for acknowledgements.
\item Use \texttt{connect} to connect to the Pedro server for acknowledgements. Pedro will be listening for the connection on the first of the two ports sent by the server.
\item Read the client ID on this socket (sent by the server as a newline terminated string).
\item Create another socket in the client, which will be used for data.
\item Use connect to connect the socket to the Pedro server for data. Pedro will be listening for the connection on the second of the two ports sent by the server.
\item Send the client the client ID on the data socket.
\item Read the status on the data socket. If the connection succeeds, the status will be the string \verb|ok\n|.
\end{enumerate}

\subsection{Kinds of Pedro messages}
\subsubsection{Notifications}
A notification is a string sent to the server ending in a newline character that represents an atom, list or compound Pedro term.\\

To process a notification, the server reads the characters in the message until it reaches a newline. Then it will attempt to parse the characters up to the newline as a compound Pedro term. If the notification parses correctly, the server will send a 1 back to the client, otherwise it will send a 0.\\

A Pedro compound term consists of an atom (the functor), an open parentheses, a comma-separated list of terms and then a closed parentheses. It is syntactically the same as a Qulog or Prolog predicate. The terms can be atoms, lists, compound Pedro terms, numbers, strings (in double quotes)\citep{Robinson2009}.\\

\subsubsection{Subscriptions}
A subscription is a special kind of notification. It is a request to receive all notifications that satisfy some condition. It always consists of a Pedro term \texttt{subscribe} with three terms, of the form:\\
\texttt{subscribe(Head, Body, Rock)}
This subscription will cause the agent to receive all notifications that unify with the \texttt{Head} of the condition and (with this unifier) satisfy the condition in \texttt{Body}.\\

\texttt{Rock} is an integer. Its meaning is determined by each client, for example it could refer to a particular message queue or a thread\citep{Robinson2009}.\\

Subscriptions can be used to request the server to only send certain notifications, for example a subscription could be sent to receive only percepts from a Pedro server. For example:
\begin{verbatim}
subscribe(controls(X), length(X)>0, 0)

\end{verbatim}

\subsubsection{Registrations}
As well as publish/subscribe, Pedro supports direct peer-to-peer communication. To do this, a client must register a name with the Pedro server. This is done by sending a \texttt{register(name)} notification, where \texttt{name} is the name being registered. The name must be an atom not containing the characters ``,'', ``:'' and ``@''. The server will acknowledge the client with a 1 if the registration succeeds and 0 if it fails\citep{Robinson2009}.\\

A registration can be removed by sending the following newline-terminated string:
\texttt{deregister(name)}
where \texttt{name} is the registered name of the process. The server will acknowledge the client with a 1 if the registration succeeds and 0 otherwise\citep{Robinson2009}.

\chapter{Language and Syntax}\label{ch:language}
\section{Implemented features}
My teleo-reactive system was mostly modelled on Qulog, but due to the scale of the task and limited amount of time in which to implement it, not all features were included\citep{Clarka,Clark2014}. The features that were included were:
\begin{itemize}
\item The ability to write TR procedures, consisting of a series of TR rules;
\item Type signatures for percepts, beliefs, actions and procedures;
\item Compile-time checking of types for percepts, beliefs, actions and procedures;
\item Dynamic modification of the BeliefStore;
\item Integration with the Pedro server, with the ability to interact with demos provided with the original Qulog distribution;
\item While/until actions.
\end{itemize}
Significant features\citep{Clarka,Clark2014} that were not included:
\begin{itemize}
\item A Prolog/Qulog-style inference system for rule conditions - it is not possible for the user to define their own predicates/relations in terms of primitive percepts and beliefs;
\item Timed sequences of actions;
\item Wait/repeat actions;
\item Moded type declarations, but these were deemed to be unnecessary because of the lack of user-defined relations;
\item Concurrency within the agent, i.e. the ability to give a teleo-reactive agent multiple tasks for it to complete simultaneously;
\item Concurrency between agents - the ability for one teleo-reactive agent to send messages / commands to another, so that they may collaborate.
\end{itemize}

\section{Syntax}
At the topmost level, the language is composed of four constructs:
\begin{itemize}
\item Type definitions;
\item Type declarations;
\item Procedure type declarations;
\item Procedure definitions.
\end{itemize}
The syntax of each of these will be described in the rest of this chapter.

\subsection{Type definitions}
Like in Qulog, type definitions can either be disjunctions of atoms, disjunction (union) of types or integer ranges. The syntax is the same as in Qulog\citep{Clark}. The following snippet shows the syntax of a disjunction of atoms type:
\begin{verbatim}
type_name ::= atom1 | atom2 | ... | atomk
\end{verbatim}
The following snippet shows the syntax of a disjunction of types type:
\begin{verbatim}
type_name ::= type1 || type2 || ... || typen
\end{verbatim}
The following snippet shows the syntax of a range type:
\begin{verbatim}
type_name ::= (min .. max)
\end{verbatim}

\subsection{Type declarations}
These have the same syntax as in Qulog\citep{Clark}, but the kinds of type declarations that can be given are restricted to \texttt{belief}, \texttt{percept}, \texttt{durative} and \texttt{discrete}. So these have the form:
\begin{verbatim}
percept_type name : type_def,
             name2 : type_def2,
             ...
             namek : type_defk
\end{verbatim}
where \texttt{percept\_type} can be either \texttt{belief}, \texttt{percept}, \texttt{durative} or \texttt{discrete}. \texttt{name} must start with a lower case letter, subsequent characters can be lower case letters, upper case letters, numerical digits or the underscore character. 

\subsection{Procedure type declarations}
These have the form:
\begin{verbatim}
procedure_name : type_def ~>
\end{verbatim}
where \texttt{procedure\_name} is a name starting with a lower case letter, where subsequent characters can be lower case letters, upper case letters, numerical digits or the underscore character. \texttt{type\_def} consists of zero or more type names inside parentheses, separated by commas, e.g. \texttt{()}, \texttt{(num)}, \texttt{(num, atom)}.

\subsection{Procedure definitions}
This is the part of the program that dictates what actions the agent performs in response to percepts and beliefs. Procedures are of the form:
\begin{verbatim}
procedure_name(params){
  rule1
  rule2
  ...
  rulek
}
\end{verbatim}
\texttt{procedure\_name} is the name of the procedure. It must have a corresponding type declaration (the syntax of which is defined above). \texttt{params} is a comma, separated list of variable names. This list of parameters must be the same length as the input types specified by the procedure type declaration. The first parameter has the first type in the declaration, and so on. \texttt{rule1} to \texttt{rulek} are the rules of the procedure.

The rules of a procedure are of the form:
\begin{verbatim}
conditions ~> actions
\end{verbatim}
where \texttt{conditions} is of one of the following forms:
\begin{verbatim}
G while WC min WT until UC min UT
G while WC until UC min UT
G while WC min WT until UC
G while WC until UC
G while WC min WT
G until UC min UT
G while WC
G until UC
G
\end{verbatim}
\texttt{G}, \texttt{WC} and \texttt{UC} are lists of conditions and \texttt{WT} and \texttt{UT} are positive numbers. If \texttt{WC} is omitted, this is the equivalent to replacing it with \texttt{true} and if \texttt{UC} is omitted, this is equivalent to replacing it with \texttt{false}. If \texttt{WT} and \texttt{UT} are omitted, this is equivalent to replacing them with \texttt{0}.

A \textbf{condition} can be either:
\begin{itemize}
\item A predicate, representing a query of a percept or belief;
\item A binary comparison of two expressions;
\item A negated condition (a condition preceded by \texttt{not}).
\end{itemize}

Every predicate representing a percept or belief must be declared as a percept or belief in a type declaration somewhere in the program. The terms of the predicate must match those in the corresponding declaration. A \textbf{predicate} consists of a name, or ``functor'' (beginning with a lower case letter or underscore, followed by upper/lower case letters, digits and underscore) optionally followed by parentheses containing one or more terms (which can be predicates, variable names or values). For example, the following strings are predicates:
\begin{verbatim}
look
see(thing, Direction, 10)
buy(hat)
sell_thing(X)
_controls(run)
_initialise
thing1
thing2(A, s23, B33)
sfs24aAfs(s, s, 55)
\end{verbatim}

The left-hand side of the rule is either a single procedure call (given as a predicate) or a comma-separated list of predicates, each referring to a primitive action to be sent to the server. A primitive action is a predicate that has been defined as a \texttt{durative} or \texttt{discrete} action in a type declaration. Every procedure call or action must refer to a declared procedure call or action. The terms of the predicate must match the declared parameters of the procedure call or action.\\

A \textbf{binary comparison} consists of two expressions, separated by a \textbf{binary comparison operator}, which can be \textbf{>}, \textbf{>=}, \textbf{==}, \textbf{<=} or  \textbf{<}. An \textbf{expression} can be an equation, which consists of two expressions separated by a binary arithmetic operator (\textbf{+}, \textbf{-}, \textbf{/} or \textbf{*}), or a single numerical value.

\chapter{System design and implementation}\label{ch:implementation}
\section{Overall design}
My language does not conform to the two-tower model\citep{Clarka}, because the ``BeliefStore tower'' in my language does not consist of inference rules that construct more abstract predicates from more ``low level'' ones. In the case of my program, the conditions on the left hand side of the rules consist only of percepts and beliefs, not relations defined in terms of them. That said, the ``Action tower'' does work at different levels of abstraction, as tasks (\textsc{TR} procedures) can be written that call other tasks (procedures), in a hierarchical fashion.\\

\section{Parser}
The first stage in the execution of a program is parsing the source file. This involves converting a flat text file into a data structure (an \textbf{abstract syntax tree}) representing the hierarchical structure of the program, according to syntactic rules. This was achieved with use of the \textit{pyparsing} parser generator library for Python\citep{pyparsing}. \textit{pyparsing} provides a domain-specific language for specifying grammars by overloading the operators \texttt{+}, \texttt{-}, \texttt{|}, etc. For example, the following Python code produces a new \texttt{ParserElement} object that, when the \texttt{.parseString} function is called, will accept the strings \texttt{>=}, \texttt{>}, \texttt{==}, \texttt{<=} and \texttt{<}. The \texttt{|} operator produces a \texttt{MatchFirst} object, which means that each of the five rules will be evaluated in sequence and the first one to parse correctly will match.\\
\begin{verbatim}
binary_comparison = Literal(">=") | Literal(">") | \
  Literal("==") | Literal("<=") | Literal("<")
\end{verbatim}

\section{Type checking}
The language performs type checking at compile-time, to ensure that percepts/beliefs/actions/procedures are all invoked with arguments of the correct type. This is to catch as many of the problems caused by misuse of types before runtime as possible\citep{typecheckinglect}. This is possible because the language requires the programmer to manually specify the types of beliefs, percepts, actions and procedures. The compiler can then scan through the defined program and check that the code conforms to these definitions. Unlike some statically typed programming languages (such as Haskell\citep{haskelltypeinference}), the compiler does not perform type inference. \\ 

\subsection{Algorithm}
Firstly, all of the type signatures of beliefs, percepts, actions and procedures are iterated over, in order to create a mapping from names of things to their corresponding types (and sorts, i.e. whether they are procedures, beliefs, etc). In doing so, duplicate definitions (and definitions that clash with built-in constructs) are caught by the parser. If a duplicate definition is detected then an error is raised and the program terminates.\\

Then the procedure definitions are checked. For each procedure, every rule is checked individually. From the type signatures, the parser can determine which predicates in the rules are beliefs, percepts, procedure calls or primitive actions, and the types of their arguments. If a predicate does not have an associated type signature, then an exception is thrown and the program terminates. Assuming that all predicates have type signatures, the parser can ensure that every procedure is typed safely (that is, every term is of the correct type and there are the correct number of terms).\\

Algorithm \ref{typecheckingalgorithm} describes the algorithm for checking that a particular object (predicate, value, etc) has is of a given type. $typeDefinitions$ is a mapping from type names to type definitions (e.g. disjunctions of atoms, disjunctions of types, range types). The mapping is determined when the program is parsed. The return statements in this algorithm of the form ``$thing$ is an X'' are boolean tests, so (for example) ``$thing$ is a predicate'' will equal True if $thing$ is a predicate.
\begin{algorithm}
\SetKwInOut{Input}{input}\SetKwInOut{Output}{output}
\Input{$thing$ - the object whose type is being checked\\
$expectedType$ - the name of the expected type of the thing\\
$typeDefinitions$ - definitions of user-defined types\\
}
\Output{$matches$ - a boolean, stating whether or not $thing$ is of $expectedType$.}

\Switch{$expectedType$}{
  \uCase{some primitive type}{
    \Return{$thing$ is a value with some primitive type}
  }
  \uCase{$predicate$}{
    \Return{$thing$ is a predicate/atom}
  }
  \Other{
    \tcc{$expectedType$ must be a user defined type}
    \eIf{$expectedType$ has a definition in $typeDefinitions$}{
      \Switch{what kind of type is $expectedType$?}{
        \uCase{disjunction of atoms}{
          \Return{$thing$ is an atom $\land$ $thing$ is one of the atoms in the type}
        }
        \uCase{disjunction of types}{
          \Return{$thing$ belongs to one of the types}
        }
        \uCase{range type}{
          \tcc{the range type has a minimum $min$ and maximum $max$ value}
          \Return{$thing$ is an $integer$ $\land$ $min \leq i \leq max$}
        }
        \Other{
          abort the program, with an ``invalue type definition'' error\;
        }
      }
    }{
      abort the program, with an ``undefined type'' error\;
    }
  }
}
\caption{Type checking algorithm \label{typecheckingalgorithm}}
\end{algorithm}

\section{Main algorithm}
The algorithm for executing the parsed program was based on the algorithm for executing \textsc{TR} programs, with some modifications made in order to accommodate the additional features. Algorithm \ref{mainloopalgorithm} describes the steps taken to initialise the program and the main perception-cognition-action loop. When the percepts have been received, Algorithm \ref{callprocedurealgorithm} is called. This algorithm takes a task (procedure) name, the program definition, the current BeliefStore, the current time, the rules that fired last time, the maximum recursion depth and the current recursion depth (which is initially 1) and returns a list of actions to be performed and the new list of fired rules. Another place where the algorithm differs to the \textsc{TR} algorithm is that a list of remembered facts is maintained - these are the beliefs that can be remembered or forgotten. The \texttt{remember/1} and \texttt{forget/1} actions add and remove beliefs from that list, respectively.\\

\begin{algorithm}
\SetKwInOut{Input}{input}\SetKwInOut{Output}{output}
\Input{$progData$ - program data\\
$procName$ - the name of the first procedure to be called\\
$rawParams$ - the parameters to pass to the procedure\\
$maxDp$ - maximum depth of recursion\\
$shellName$ - the shell name\\
$serverName$ - the server name}

Initialise the Pedro client, with the $shellName$ and $serverName$ parameters\;

$parsedParams \leftarrow parse(rawParams)$\;

$startTime \leftarrow $ the current system time\;

$currentTime \leftarrow 0$\;

$LastActions \leftarrow []$\;

$rememberedFacts \leftarrow []$\;

$previously fired rules \leftarrow []$\;

\While{$True$}{
  $t \leftarrow $ the current system time\;

  $currentTime \leftarrow t - currentTime$\;

  $percepts \leftarrow$ percepts from the percept source (i.e. Pedro)\;

  $beliefStore \leftarrow percepts + rememberedFacts$\;

  Call the top task call, taking the following input parameters:
    the previously fired rules
    the belief store
    parsed procedure call parameters
    the procedure name
    the program data
    current time\;
  This will return:
    a list of actions to perform
    a list of fired rules\;

  $primitiveActions \leftarrow []$\;

  \For{every action in the list}{
    \Switch{what kind of action is it?}{
      \Case{a remember(X) action}{
        add the belief X to $rememberedFacts$\;
      }
      \Case{a forget(X) action}{
        remove the belief X from $rememberedFacts$, if present\;
      }
      \Other{
        add it to $primitiveActions$\;
      }
    }
  }

  Send every action in $primitiveActions$ to the Pedro agent, using the initialised Pedro client\;
}
\caption{MainLoop\label{mainloopalgorithm}}
\end{algorithm}

\subsection{Procedure call}
This part of the program is called once by the main algorithm for every iteration of the main loop. It returns the actions the agent should perform / send to Pedro and a list of all of the fired rules. Every time it is called, it checks if the call depth limit has been exceeded. If so, it aborts with an error. Otherwise, it evaluates the rules in the current procedure to find which rule should fire.\\

The algorithm returns a list of the currently fired rules, so that when it evaluates the procedure it is able to execute the while/until semantics (see Algorithm \ref{getactionalgorithm}). If a rule at a certain depth of recursion stops firing, then all of the rules lower down in the call hierarchy (i.e. ``child'' rules called by it) are removed from the list of currently fired rules. This is the same as the behaviour of $FrdRules$ in step 11 and 12 of the the \textsc{TR} algorithm, shown in Figure \ref{fig:flowchart} in Chapter \ref{ch:background}.\\

\begin{algorithm}
\SetKwInOut{Input}{input}\SetKwInOut{Output}{output}
\Input{$progData$ - the program data\\
$procName$ - the name of the procedure to be called\\
$parsedParams$ - procedure call parameters\\
$beliefStore$ - belief store\\
$currentTime$ - current time\\
$previousRules$ - list of previously fired rules\\
$maxDp$ - max depth of recursion\\
$Dp$ - current depth of recursion}

\Output{$actions$ - list of the actions to be performed\\
$firedRules$ - a list of currently fired rules}

\If{$Dp > maxDp$}{
  abort the program, with an error "exceeded-recursion-depth"\;
}

look up the current procedure called $procName$ in $progData$\;

$variables \leftarrow$ the arguments that were passed to the procedure, mapped to the names used inside the procedure\;

\If{$length(previousRules) > Dp$}{
  \tcc{then a rule at the same depth previously fired}
  $previousRule \leftarrow$ the last rule at the same depth, that previously fired\;
  $newFiring \leftarrow GetAction(beliefStore, rules, variables, currentTime, previousRule)$\;

  replace $previousRule$ in $previousRules$ with $newFiring$ to get $newPreviousRules$\;
  \If{$newFiring$ is for a different rule or different actions to $previousRule$}{
    delete all subsequent rule firings in $newPreviousRules$\;
  }
}
\Else{
  \tcc{no rule at the same depth previously fired}
  $newFiring \leftarrow GetAction(beliefStore, rules, variables, currentTime, null)$\;
  append $newFiring$ to $previousRules$ to get $newPreviousRules$\;
}

\Switch{What are the new actions?}{
\Case{a procedure call}{
  call the procedure using this algorithm (recursive call): $CallProcedure(progData, newProcName, newParsedParams, beliefStore,$\\
  $currentTime, newPreviousRules, maxDp, Dp + 1)$\;
  \Return{the result of calling the procedure}\;
}

\Case{a tuple of primitive actions}{
  \Return{the tuple of actions and the list of current rule firings}\;
}
}
\caption{CallProcedure \label{callprocedurealgorithm}}
\end{algorithm}

\subsection{Get Action}
Algorithm \ref{getactionalgorithm} returns the next rule firing for a procedure, given the belief store, the procedure definition, the instantiated variables in the procedure call, the current time and the previous rule firing. The previous rule firing and current time are needed to execute the semantics of the while/until conditions. If a rule previously fired in this procedure with the same instantiated variables, then the while/until continuation conditions can be evaluated. Otherwise, the algorithm evaluates the guard conditions in the same way as the \textsc{TR} algorithm. The ``firing'' object keeps track of the time the rule was first fired, because this is what is used to determine whether the minimum time limits for while and until have expired.

\begin{algorithm}
\SetKwInOut{Input}{input}\SetKwInOut{Output}{output}
\Input{
$beliefStore$ - belief store\\
$procedure$ - list of rules for the given procedure\\
$variables$ - variable instantiation for this call\\
$currentTime$ - current time\\
$prevFiring$ - the previous rule firing (this can be null)
}

\Output{
$newFiring$ - the new rule firing
}

$ruleFound \leftarrow False$\;

\If{$prevFiring$ is not null}{
  look up the rule in the procedure that fired previously\;
  $WC \leftarrow$ the rule's while condition\;
  $WT \leftarrow$ the rule's while minimum time limit\;
  \If{
    $WC$ is inferable under $beliefStore$, with the instantiated $variables$
    $\lor$
    $WT$ has not expired since the rule's first firing
  }{
    $UC \leftarrow$ the rule's until condition\;
    $UT \leftarrow$ the rule's until minimum time limit\;
    \If{$UC$ is NOT inferable under $beliefStore$ with the instantiated $variables$
      $\lor$
      $UT$ has not expired since the rule's first firing}{

      $ruleFound \leftarrow True$\;
      $newVariables \leftarrow variables$\;
      $firstFired \leftarrow$ whenever $prevFiring$ first fired\;
      set the output to be the current rule firing\;
    }
  }
}

\If{$ruleFound = False$}{
  starting with the first rule in the procedure,
   find the first rule that has inferable guard conditions under the current belief store\;

  \If{a rule is found}{
    $ruleFound \leftarrow True$\;
    $R \leftarrow$ the number of the rule (i.e. the first rule in the procedure will be 0, etc)\;
    $newVariables \leftarrow$ the variable mapping from the procedure parameters, plus variables instantiated when the guard was evaluated\;
    $firstFired \leftarrow currentTime$\;
  }
}

\If{$ruleFound = True$}{
  \Return{\{ 'actions' : the actions on the RHS of the fired rule,
      'R' : the ID of the rule,
      'first\_fired' : $firstFired$
      'variables' : $newVariables$ \}
  }\;
}
\Else{
  abort with an error "no-firable-rule"\;
}
\caption{GetAction \label{getactionalgorithm}}
\end{algorithm}

\section{Evaluation of conditions}
On the left-hand side of the rules, a list of conditions is given. A condition can be a belief/percept query (e.g. \texttt{see(thing, left, Distance)}), a negated belief/percept query (e.g. \texttt{not see(thing, right, 10)}) or a binary comparison (e.g. \textsc{X > 2}).\\

The evaluation of conditions can be split into two parts:
\begin{itemize}
\item Evaluating a single individual condition;
\item Evaluating a conjunction of conditions.
\end{itemize}
Firstly, the way in which a single condition is evaluated will be explained. Then, it will be explained how a conjunction of conditions is evaluated.\\

The algorithm to evaluate a single condition (Algorithm \ref{conditionsalgorithm}) takes three inputs: $condition$ the condition itself, $beliefStore$ the current set of beliefs/percepts and $variables$ a mapping from variable names to values. The algorithm returns a boolean $success$ which is True if the condition can be inferred and $bindings$ which is a list of all possible mappings of variables under which the condition is inferable (where every mapping contains all the mappings in $variables$).\\

\begin{algorithm}
\SetKwInOut{Input}{input}\SetKwInOut{Output}{output}
\Input{
$condition$ - the condition being evaluated, which can be a belief\/percept query, a negated belief\/percept query or a binary comparison\\
$beliefStore$ - the BeliefStore. This is a list of beliefs and percepts known by the agent\\
$variables$ - the existing mapping of variable names to instantiated values\\
}
\Output{
$success$ - a boolean, this is true if the condition is true for some variable mapping\\
$bindings$ - a list of all of the possible mappings from variables to values, that include the mappings in $variables$\\
}
\Switch{what is $condition$?}{

\uCase{$condition$ is a binary condition}{
  evaluate the binary condition, with respect to the instantiated variables\;
  \If{the binary condition is satisfied}{
    $success \leftarrow True$\;
    $bindings \leftarrow$ singleton list containing $variables$\;
  }
}
\uCase{$condition$ is a negated query}{
  evaluate the original query\;
  \eIf{the query succeeds}{
    $success \leftarrow False$\;
    $bindings \leftarrow None$\;
  }{
    $success \leftarrow True$\;
    $bindings \leftarrow$ singleton list containing $variables$\;
  }
}
\uCase{$condition$ is a query}{
  find all of the variable instantiations that cause $condition$
    to match with some predicate in $beliefStore$
    where all variables in $condition$ have been replaced
    with the associated values in $variables$ (if possible)\;

  \eIf{$cond$ cannot match with anything in $beliefStore$}{
    $success \leftarrow False$\;
    $bindings \leftarrow None$\;
  }{
    $success \leftarrow True$\;
    $bindings \leftarrow$ a list of all the returned variable instantiations\;
  }
}
\Other{
  Abort with error: invalid condition\;
}
\caption{Algorithm for evaluating conditions \label{conditionsalgorithm}}
}

\Return{success, bindings}
\end{algorithm}

\subsection{Evaluation of a single query}
This part will describe the process for evaluating whether a single query (a predicate) is inferable. This is similar to answering the question ``is this predicate present in the BeliefStore?'', where the BeliefStore is a list of ground predicates, but is made slightly more complicated by the fact that query conditions can have variables as their terms. So instead the algorithm checks if some ground predicate in the BeliefStore is able to ``match'' with it. For a ground predicate to ``match'' with a query, there must be some mapping of variables to ground values (or, ``instantiation''), that when applied to the query causes it to be \textbf{syntactically equal} to the ground predicate. Two arguments will be returned by this algorithm: a boolean saying whether the query is inferable and (if the query is inferable) a list containing all the instantiations or (if the query is not inferable) a null value.\\

The ``match'' condition is less sophisticated than unification\citep{martelli1976unification}, because only one argument will ever contain any variables (the query condition). Contrast this with unification, where both objects can be or contain uninstantiated variables. This restriction also makes the use of moded type declarations unnecessary because a query to the BeliefStore will ground all variables in the query.\\

Algorithm \ref{queryalgorithm} describes the process of evaluating a single query and Algorithm \ref{patternmatchalgorithm} describes the process for matching a query (with variables) with a ground predicate.\\

\begin{algorithm}
\SetKwInOut{Input}{input}\SetKwInOut{Output}{output}
\Input{
$condition$ - the query to be evaluated\\
$beliefStore$ - a list of predicates, the BeliefStore\\
$variables$ - the current variable mapping
}
\Output{
$success$ - boolean, True if $condition$ matches with something in $beliefStore$\\
$bindings$ - list of all variable mappings that cause $condition$ to match with\\something in $beliefStore$, that include the mappings already in $variables$
}
$bindings \leftarrow []$\;
\For{$fact \in beliefStore$}{
  try to match $fact$ with $condition$, given the existing variable instantiation $variables$\;
  \If{$fact$ can match with $condition$}{
    add its variable instantiation to $bindings$\;
  }
}
\eIf{length of $bindings$ $=$ $0$}{
  \Return{False, []}\;
}{
  \Return{True, $bindings$}\;
}
\caption{Algorithm for evaluating a single query \label{queryalgorithm}}
\end{algorithm}

\begin{algorithm}
\SetKwInOut{Input}{input}\SetKwInOut{Output}{output}
\Input{
$ipred$ - input predicate\\
$gped$ - ground predicate\\
$vars$ - variable mapping\\
}
\Output{
$success$ - boool, True if matches\\
$newvars$ - new variable inst!\\
}
\uIf{ipred is a variable}{
  \eIf{$ipred \in vars$}{
    $p \leftarrow$ look up ipred in vars\;
    $s$, $newvars$ $\leftarrow$ match(p, $gpred$, $vars$)\;
    \Return{$s$, $newvars$}\;
  }{
    $newvars \leftarrow vars$\;
    add $ipred \rightarrow gpred$ to $newvars$\;
    \Return{True, $newvars$}
  }
}
\uElseIf{ipred is a predicate}{
  $vars\_temp \leftarrow vars$\;
  $s1 \leftarrow T$\;
  $ipredArgs \leftarrow$ the terms in $ipred$\;
  $gpredArgs \leftarrow$ the terms in $gpred$\;
  \eIf{length($ipredArgs$) $=$ length($gpredArgs$)}{
    \For{$x$,$y$ $\in$ zip(, gpred.args)}{
      \tcc{try to match each of the pairs}
      $s1$, $vars\_temp$ $\leftarrow$ match($x$,$y$,$vars\_temp$)\;
      \If{$s1$ is False}{
        \tcc{two terms do not match}
        \Return{False, null}
      }
    }
  }{
    \tcc{the query and predicate have different number of terms, definitely do not match}
    \Return{False, null}
  }

  \Return{True, $vars\_temp$}
}
\Else{
  throw error $ipred$ is unrecognised\;
}
\caption{Algorithm for matching a query with a ground predicate \label{patternmatchalgorithm}}
\end{algorithm}

\subsection{Evaluation of binary comparisons}
A binary condition is evaluated by initialising any variables on both sides of the comparison using the mapping in \texttt{vars}, evaluating the arithmetic expressions on both sides, then comparing the resulting values. The possible comparisons are:
\begin{itemize}
\item \texttt{>} - greater than;
\item \texttt{>=} - greater than or equal;
\item \texttt{==} - equal;
\item \texttt{<=} - less than or equal;
\item \texttt{<} - less than.
\end{itemize}
For a binary comparison to take place, the expressions on either side must have evaluated to ground numerical values. So it is not possible to evaluate something like \texttt{X > 3} when \texttt{X} has not been instantiated, this will result in an error being thrown.

\subsection{Evaluation of arithmetic expressions}
An arithmetic expression is evaluated recursively. The order of precedence of operators is handled by the parser. It produces a binary tree, which is traversed every time the expression is evaluated. Depending on what the expression is, it can be evaluated in one of three ways:
\begin{itemize}
\item a ground numerical value, i.e. an integer or floating point number. Return the numerical value;
\item a single variable, e.g. \texttt{X}. Look up the value in the variable mapping, then return it. If there is no value then raise an error, as the expression cannot be evaluated.
\item two arithmetic expressions and a binary operator such as \texttt{+}, \texttt{-}, \texttt{/} or \texttt{*}. Evaluate the left and right hand expressions, then apply the relevant operator.
\end{itemize}

\chapter{Experiments \& Evaluation}\label{ch:evaluation}
\section{Demonstration}
One way to demonstrate that the teleo-reactive system I have produced works is to have it interact with something. This section outlines how that can be done and gives an example demonstration.

\subsection{Creating a demonstration}
As mentioned before, a teleo-reactive program takes a stream of percepts as input and produces a stream of actions. Therefore, in order to demonstrate a teleo-reactive program, it is necessary to provide a program that accepts a stream of actions and produces a stream of percepts. For example, this could be a game (if one wanted to create a video game playing ``bot''), a simulation of a real world robotics scenario or an interface to a real robot's actuators and sensors. Intermediate layers between the actuators, sensors and the teleo-reactive program can also be incorporated, for example one could perform object recognition on a video input.\\

In the case of Qulog and my system, communication of percepts and actions between the teleo-reactive program and the environment is handled by the Pedro communication system, which is explained more in previous sections. The Pedro distribution comes with C, Java and Python APIs so it is possible to write a program that interacts with teleo-reactive systems without understanding the communications protocol itself \citep{Robinson2009}.

\subsection{Asteroids demonstration}
\begin{figure}[ht!]
\centering
\fbox{
\includegraphics[width=90mm]{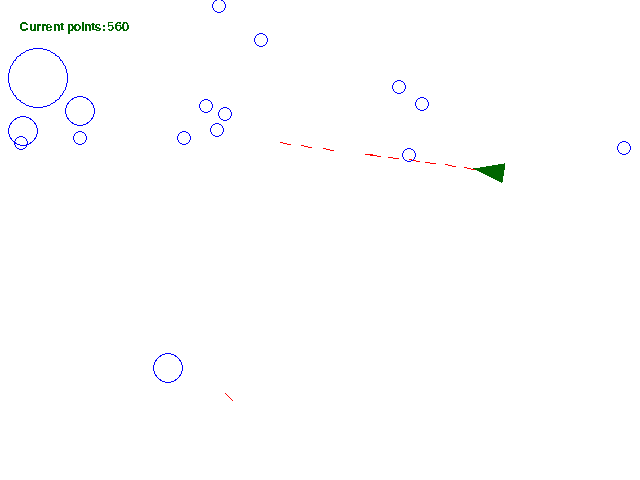}
}
\caption{Screenshot of the asteroids game \label{fig:asteroidsscreenshot1}}
\end{figure}

I adapted an asteroids game I had previously programmed (in Python, using the Pygame library (\url{http://www.pygame.org}) to send and receive Pedro percepts\citep{asteroidsBob2015}. The details of how the game itself was implemented are not relevant to this report. Figure \ref{fig:asteroidsscreenshot1} shows a screenshot of a game in progress.\\

The game itself involves a spaceship that can move around a 2D world and shoot bullets. Asteroids (depicted as circles) float around the world. The objective is to control the spaceship to shoot all of the asteroids, without being hit by an asteroid.\\

The game produces three kinds of percepts:
\begin{itemize}
\item \texttt{see(Obj,Dir,Dist)} - ``the spaceship sees the object \texttt{Obj} in the direction \texttt{Dir} (left, right, centre or dead\_centre) at a distance of \texttt{Dist} (pixels)'';
\item \texttt{facing\_direction(RDir)} - ``the spaceship is currently facing in the direction \texttt{RDir} (measured in radians, where $0^r$ is east)'';
\item \texttt{speed(S)} - ``the spaceship is currently travelling with speed S''.
\end{itemize}

One \texttt{see/3} percept is produced for every single thing the spaceship sees. In this game, the only thing that can be seen are asteroids. The number of \texttt{see/3} percepts that are generated depends on how many asteroids are in the spaceship's field of vision. It expected that a real robotic agent will not know the full state of the world, so it makes sense to restrict a simulated model's knowledge as well, to demonstrate how teleo-reactive programming can operate on incomplete sensory information.\\

\begin{figure}[h!]
\centering
\includegraphics[trim=3cm 20cm 6cm 0cm,clip]{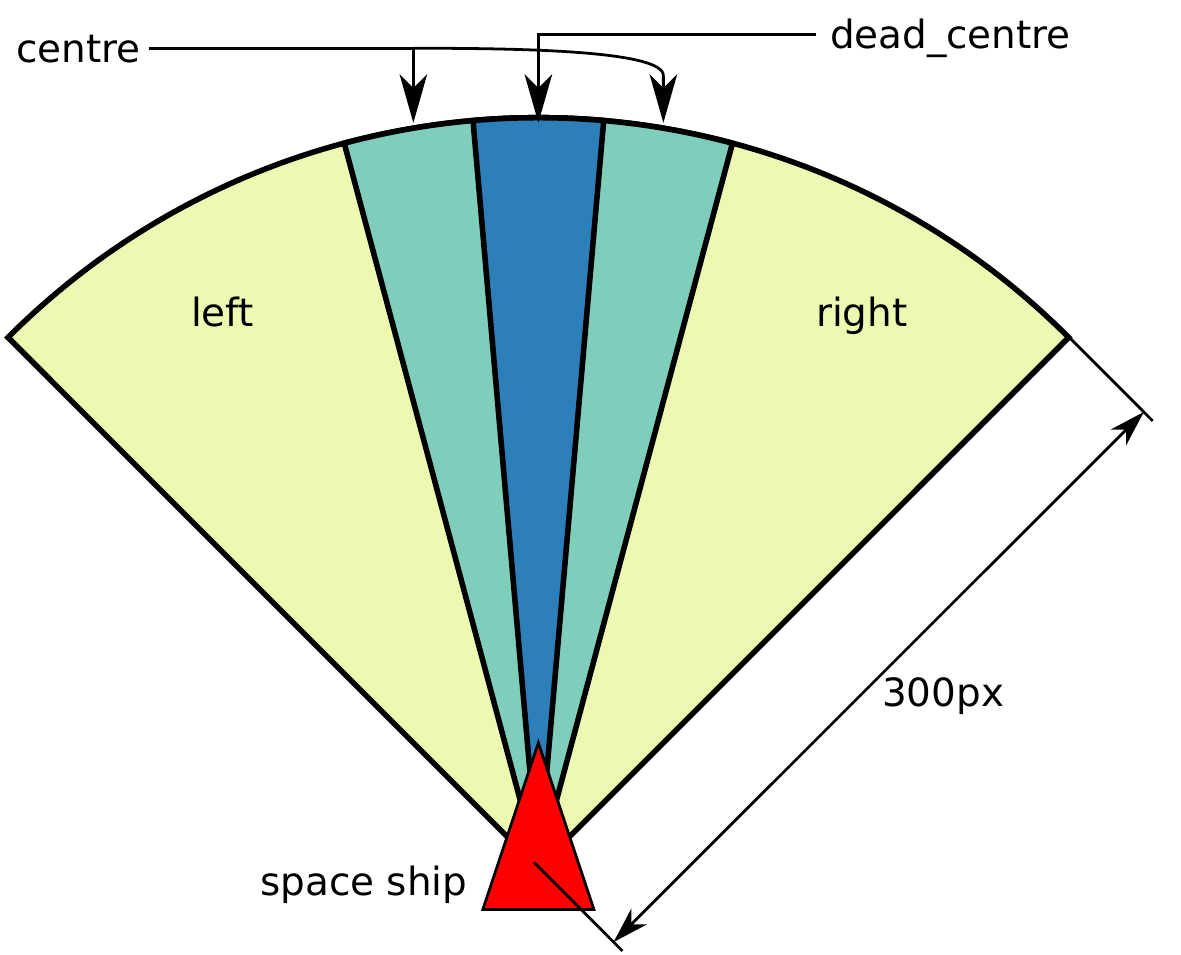}
\caption{Diagram showing the behaviour of the \texttt{see/3} percept}
\label{fig:seeperceptdiagram}
\end{figure}

If the asteroids are almost directly in front of the space ship, then the \texttt{Dir} term of the percept will be \texttt{dead\_centre}. If the asteroids are almost in front of the space ship, but not close enough to be \texttt{dead\_centre}, the term will be \texttt{center}. If they are to the left or right (up to some angle) then the term will be \texttt{left} or \texttt{right}. So asteroids that are behind the spaceship will not be sensed. Asteroids that are more than 300 pixels away from the asteroid will also not be sensed. The space ship's field of vision is illustrated in the diagram (Figure \ref{fig:seeperceptdiagram}).\\

The actions that can be received by the program are:
\begin{itemize}
\item \texttt{move\_forward} - move the space ship forward;
\item \texttt{move\_backward} - move the space ship backward;
\item \texttt{turn\_left} - turn the space ship left;
\item \texttt{turn\_right} - turn the space ship right;
\item \texttt{shoot} - shoot.
\end{itemize}

All of these actions are durative, which means that one signal is sent to Pedro to indicate that the action is to start, then another one is sent when the action ends. This is to be contrasted with discrete actions, where one signal is sent to tell the agent to do the action.\\

\subsubsection{Example programs}
Every program for this demonstration has the same type definitions and declarations - these define the valid inputs and outputs. These are:\\

\begin{verbatim}
thing ::= asteroid | something_else
direction ::= left | right | centre | dead_centre

durative turn_right : (),
         turn_left : (),
         move_forward : (),
         move_backward : (),
         nothing : (),
         shoot : ()

percept facing_direction : (num),
        see : (thing, direction, num),
        speed : (num)
\end{verbatim}
This defines two types: \texttt{thing} and \texttt{direction}. It also declares the types of six durative actions and three percepts. These are then referred to in the right hand side and left hand side of the rules, respectively. Procedures can now be written in terms of these percepts and actions.\\

The simplest procedure consists of one rule with a guard that is always inferable (\texttt{true}), which tells the agent (space ship) to do nothing. The procedure has no arguments and so has a type of \texttt{()}. Its is written:
\begin{verbatim}
proc1 : () ~>
proc1(){
true ~> ()
}
\end{verbatim}

If the agent is to do anything, an action will have to be placed on the right hand side. This procedure tells the space ship to move forward, indefinitely:
\begin{verbatim}
proc2 : () ~>
proc2(){
true ~> move_forward
}
\end{verbatim}

Now for a more advanced program. The verbose nature of this code was a result of a feature being accidentally missed out - the ability to put a conjunction of queries under a negation by failure statement (e.g. ``q(a) \& not (b(X) \& c(X))''). \texttt{inner\_proc\_left} and \texttt{inner\_proc\_right} function as negation by failure statements.\\

\begin{verbatim}
proc3 : () ~>
proc3(){
see(asteroid, left, D) ~> inner_proc(D)
see(asteroid, right, D) ~> inner_proc(D)
true ~> move_forward
}

inner_proc : (num) ~>
inner_proc(D){
see(asteroid, left, D2) & D2 < D ~> inner_proc(D2)
see(asteroid, right, D2) & D2 < D ~> inner_proc(D2)
see(asteroid, left, D2) & D2 == D ~> turn_left, shoot
see(asteroid, right, D2) & D2 == D ~> turn_right, shoot
}
\end{verbatim}
In this program, \texttt{proc3} calls \texttt{inner\_proc} with a numerical argument \texttt{D} which represents the current distance asteroid that has just been spotted (that is on the left or right of the space ship). \texttt{inner\_proc} recursively calls itself until it has found the closest asteroid to the space ship (that is on the left or right of the space ship). Then, depending on the side, it tells the space ship to turn left or right and shoot. This is definitely not the best way to write this program, the intended meaning of the program was more like:\\
\begin{verbatim}
proc3 : () ~>
proc3(){
see(asteroid, left, Dist1) & 
  not (see(asteroid,X,Dist2) & Dist2 < Dist1) ~> turn_left, shoot
see(asteroid, right, Dist1) &
  not (see(asteroid,X,Dist2) & Dist2 < Dist1) ~> turn_right, shoot
true ~> move_forward
}
\end{verbatim}

This program is able to play Asteroids and win some games. It prioritises destroying the nearest asteroid. However, it sometimes exhibits strange behaviour because it does not take all information about the world into account. If an asteroid is travelling near the space ship and is about to pass it, the space ship may not be able to turn round quickly enough to aim at it. This sometimes causes it to miss.\\

Another situation that confounds this program is when there are multiple asteroids all at a similar distance from the space ship, the space ship will constantly be changing direction to aim at a different one. The result being that it does not succeed in aiming at any of them. One way to address this could be to set a small minimum limit on the amount of time that a rule can fire. This could be done by adding a ``while min T'' clause to the rule conditions, where \texttt{T} is a positive floating point number. No while condition is given, so it defaults to \texttt{false}. So the revised rules would look like  this:\\
\begin{verbatim}
proc3 : () ~>
proc3(){
see(asteroid, left, Dist1) &
  not (see(asteroid,X,Dist2) &
  Dist2 < Dist1) while min 0.1 ~> inner_proc_left(Dist1)

see(asteroid, right, Dist1) &
  not (see(asteroid,X,Dist2) &
  Dist2 < Dist1) while min 0.1~> inner_proc_right(Dist1)
true ~> move_forward
}
\end{verbatim}

Another program that could be written is one that maintains some percept, such as the speed of the space ship. The following program contains a procedure \texttt{regulate\_speed} which takes a number as input and ensures that the space ship travels at that speed. It does so by speeding up when its speed drops below the level and doing nothing otherwise (the game world has some friction so the ship is inclined to slow down gradually). This example shows how self-regulating behaviour can be programmed.\\
\begin{verbatim}
proc4 : () ~>
proc4(){
true ~> regulate_speed(5)
}

regulate_speed : (num) ~>
regulate_speed(D){
speed(S) & S < D ~> move_forward
true ~> ()
}
\end{verbatim}

\subsection{Evaluation of the demonstration}
The problem that I set out to investigate with teleo-reactive programming was that of developing robust, opportunistic programs to control robots in continuously varying environments. In this situation, a program is robust if it is able to recover from failures or setbacks. A program is opportunistic if it is able to take advantage of fortunate circumstances to more quickly achieve its goals. I do not believe that the Asteroids demo was sufficient to demonstrate that teleo-reactive programming was robust or opportunistic for these reasons:\\

To win at asteroids, the player has to do two things: shoot at asteroids and avoid being hit by asteroids. Unlike solving a puzzle, or performing a complex task, these tasks are not (obviously) made up of a high number of sub-tasks. If an agent is pursuing some strategy that is made up of one or two steps, if something happens to disrupt its behaviour, it will be set back at most one or two steps. This kind of robustness is simple enough that it can be hand-coded into a non-teleo-reactive program. In short, the Asteroids problem does not require a sufficiently \textbf{hierarchical} solution for it to be a good test of teleo-reactive programming.\\

That said, the Asteroids demo does demonstrate how easy it is to program applications that continuously react to the environment using teleo-reactive programming. For example, a procedure can be written in about eight lines (in a language that lets the programmer write negation of failure of conjunctions of predicates) that can play a game of Asteroids. Still, in order for this to be properly tested, a more complex example would need to be constructed.\\

\section{Evaluation of my implementation}
From the perspective of the user, my implementation of teleo-reactive programming has far fewer features than Qulog/\textsc{TeleoR}. Features that it lacks\citep{Clark2014, Clarka} include:
\begin{itemize}
\item The ability for the user to define their own functions and relations;
\item The ability to execute tasks (procedures) in parallel;
\item Communication with other teleo-reactive agents - communication is solely between the agent and its sensors / actuators;
\item Timed sequences of action;
\item Wait/repeat actions;
\item Moded type declarations;
\end{itemize}
However, this does not entirely exclude it from being of use for simple tasks. Situations where feedback loops or reactive behaviour are required can be modelled in this application (as shown by the Asteroids example).

\section{Practical considerations}
This section concerns issues about the project that do not specifically correspond to teleo-reactive programming.

\subsection{Why was Python used?}
The \textsc{TR} algorithm stated in \citep{Clarka, Clark2014} formed the initial basis of the teleo-reactive system. I chose Python because it is the programming language I am most comfortable with and the fact that it is typically used as an imperative language meant that the \textsc{TR} algorithm easily translated to Python code. The choice of language was mostly a matter of personal preference.\\

Initially, I considered embedding Qulog in Haskell because simpler logic programming languages had already been successfully embedded in it\citep{Claessen2000, Spivey1999}. This course of action was not taken in the end because it was agreed that it might not be feasible to embed enough of Qulog in Haskell to produce a complete working system in the time available.\\

\subsection{Can the teleo-reactive system be compiled?}
Python programs are typically run as interpreted programs, but can be converted to stand-alone executables that do not require Python to be installed to run. This can be done using a programs such as Pyinstaller\citep{pyinstaller2015} or py2exe\citep{py2exe2015}. The compiled program can then run on the computer controlling the robot, providing that it can run one of the operating systems supported by Pyinstaller/py2exe.\\

\subsection{How does it scale?}
No empirical analysis was done of the system's scalability (i.e. by testing it on a very large input). With that said, some points where the program could be made more efficent were identified.\\

Redundancy when evaluating guard conditions - every time a single guard condition is queried, every predicate(percept/belief) in the BeliefStore is checked against it. Depending on the number of sensors, the amount of predicates (i.e. sensor data) in the BeliefStore could be very high. One solution is to use a relational database like MySQL or SQLite\citep{mysql1995mysql, hipp2007sqlite}. Every percept and belief could be represented by a table in the database. All of the variable instantiations for a single guard condition can be returned by a single SQL query.\\

Modifications to the teleo-reactive component of the system should also be considered. The optimisation described in Section \ref{subsec:continuation} could be used to minimise the number of times that rules are re-evaluated, by only querying the relevant guard conditions. Another optimisation to consider is that of Mousavi and Broda \citep{mousavi2003simplification}, which offers methods for simplfying teleo-reactive sequences by removing redundant rules and literals.\\

\subsection{Why did you write an interpreter?}
The objective of this project was to explore teleo-reactive programming, rather than one into compiler design. Interpreters are easier to write than compilers because they do not involve a code generation step. Since code generation was not the aim of this project, the decision to write an interpreter was taken.\\

Another alternative to an interpreter is an embedding of the language. A ``shallow embedding'' of Qulog in Haskell was originally considered, along the lines of\citep{Claessen2000,Spivey1999}. A shallow embedding is when the operations of the language (i.e. Qulog) are implemented in the host language (i.e. Python or Haskell). In the case of Qulog, this would be functions to perform logic programming (unification, resolution, search) for the guard conditions, evaluate \textsc{TR} procedures, functions and concurrency. There was no reason other than familiarity with the idea of writing interpreters and compilers that an interpreter was developed and not an embedding.\\

\section{Other implementations of teleo-reactive programming}
Many variations and extensions of the original \textsc{TR} language \citep{Nilsson1994} have been developed over the years. A systematic review of teleo-reactive programming produced in 2014 provided references to most of the following papers\citep{morales2014systematic}. The first one cited by \citep{morales2014systematic} is \citep{lee1994structured} which proposes a new circuit semantics \textit{Structured Circuit Semantics}. This addresses some shortcomings with Nilsson's original formalism:
\begin{itemize}
\item \textbf{Execution Cycle} - The paper proposes that because a teleo-reactive system as implemented on a computer works (at some level) in a discrete fashion, then durative (here called ``energized'') actions should instead be treated as many discrete (or ``ballistic'') actions. This makes no difference to the actual operation of the agent, so long as the perception-cognition-action frequency (i.e. the frequency of the main \textsc{TR} loop) is at or higher than the characteristic frequency of the system the agent is operating in.
\item \textbf{Non-Deterministic Behaviour} - the original TR language requires the programmer to order the rules in a procedure strictly and statically (i.e. they must be in some order and the order cannot change mid-execution). Lee and Durfee's modification allows the programmer to specify multiple ``equally good'' actions using a \textbf{do any} statement. This takes multiple rules and tries one non-deterministically. If it fails, it will try another one. If every rule fails then the whole rule fails.
\item \textbf{Best-First Behaviour} - Sometimes, the best rules to consider (i.e, the order in which rules should be given) depends on some dynamically changing \textit{cost function}. The modified language includes a \textbf{do best} statement that can contain multiple rules, each with a cost function. When such a statement is evaluated, the cost functions of each rule are evaluated and the condition/action pair of the one with the best cost function is then considered. This introduces an extra level of reasoning, which the paper called the ``decision layer''.
\item \textbf{Failure Semantics} - In SCS, every primitive action returns ``success'' or ``failure'' depending on whether it successfully had an effect on the world. These propagate upwards, for example if a group of actions are to be performed sequentially, then that group of actions will fail if any one of the actions fails. This success/fail message determines how \textbf{do any} and \textbf{do best} behave.
\end{itemize}
In addition to the above features, there are a few others such as parallel execution of steps, the \textbf{do first} structure (which chooses the first rule whose conditions are satisfied, much like in a TR program)\citep{lee1994structured}. The syntax is also different to \textsc{TR} \citep{Nilsson1994} and \textsc{TeleoR} \citep{Clarka}.\\ 

In 1995, Zelek introduces TR+ , an extension to \textsc{TR} which lets actions be performed concurrently. The programmed agent was also able to perform dynamic real time path planning. TR+ was tested out on real life robots and software tools were developed to aid the development of TR+ applications: PVM, which allows TR+ programs to be run on a cluster of heterogenous computers (i.e. the computers do not have to be identical) and a program for displaying and manipulating TR+ programs as tree graphics\citep{zelek1995teleo}.\\

Nils Nilsson and Scott Sherwood Benson extended \textsc{TR} in \citep{benson1993reacting} to allow the agent to work towards achieving multiple goals with different and time-varying urgencies, take advantage of a planning system to augment its own program and to change its program based on input from humans, successful/unsuccessful experiences and supervised/unsupervised training methods\citep{benson1993reacting}.\\

In 1996, Scott Sherwood Benson describes \textsc{TRAIL} (Teleo-Reactive Agent with Inductive Learning). It does so by recording its environment as it executes a plan or by observing a trainer. Then it identifies instances of action success or failure. Finally it uses a variant of the Inductive Logic Programming algorithm DINUS to induce action models (TR programs) based on the recorded data. This is applied to several problems domains: a 2D ``Botworld'' which involves a robot grabbing a bar, the ``Delivery domain'' where a robot has to navigate between rooms to perform some task and a flight simulator, where \textsc{TRAIL} was connected to an existing flight simulator\citep{benson1996learning}.\\

In 2003, Mousavi and Broda describe a method for simplfying TR sequences. An algorithm is presented which, given a TR program, returns one that is smaller but semantically equal to the original. It does so by removing redundant rules (rules which will never be evaluated) and redundant literals (predicates that do not affect the result of the computation). It also shows how this method can be applied to simplify decision lists\citep{mousavi2003simplification}.\\

In 2008, Gubisch et al introduce an architecture based on teleo-reactive programming for control of mobile robots. This implementation translates the given \textsc{TR} program into C++ code, which can then be compiled by the \textsc{g++} C++ compiler into a program that can be executed quickly by the robot. The paper describes how this was applied to control RoboCup (a robotic soccer competition) robots\citep{gubisch2008teleo}.\\

In 2014, Soto, S{\'a}nchez, Mateo, Alonso and Navarro developed an educational tool that converts teleo-reactive programs to VHDL, VHSIC Hardware Description Language. This language can then be used in programmable hardware devices, such as field programmable gate arrays (FPGAs)\citep{soto2014educational}.\\

In 2012, Robert Kowalski and Fariba Sadri gave a semantics for teleo-reactive programs, expressed in terms of abductive logic programming (ALP). It compares the semantics to that of LPS (Logic-based Production System and agent language), another language that can also be expressed in terms of ALP\citep{kowalski2012teleo}.\\

In summary, in addition to \textsc{TeleoR}\citep{Clark2014,Clarka}, various other extensions of the original \textsc{TR} language have been proposed \citep{lee1994structured, zelek1995teleo, benson1993reacting, benson1996learning}. Various methods of implementing teleo-reactive programs have also been introduced, such as compiled \textsc{C++} code\citep{gubisch2008teleo}, virtual hardware descriptions\citep{soto2014educational} and abductive learning\citep{kowalski2012teleo}.\\

\section{Alternatives to teleo-reactive programming}
According to the paper \citep{dongol2014reasoning}, the advantage that teleo-reactive programming provides over formalisms such as continuous action systems\citep{back2000generalizing, meinicke2006continuous}, TLA+\citep{lamport2002specifying} and hybrid automata\citep{henzinger2000theory} is that it uses durative actions that specify a behaviour over time instead of instantaneous discrete actions. This and that fact that teleo-reactive programs have a hierarchical organisation means that the representations can be less complicated than the equivalent action system\citep{dongol2014reasoning}.\\

One system that could perform a similar role to teleo-reactive programming is functional reactive programming (FRP). It can be used to describe hybrid systems - which are systems that combine both continuous and discrete components\citep{hudak2003arrows}. FRP extends functional programming to allow programs where the outputs continuously react to their inputs to be written\citep{nilsson2002functional}. Several systems exist to allow this, such as Elm \citep{czaplicki2012elm} (used for creating web-based graphical user interfaces) and many Haskell libraries \citep{frphaskellwiki}.\\

\chapter{Conclusion \& Future Work}\label{ch:conclusion}
This report has given a detailed background summary on two instances of teleo-reactive programming, \textsc{TR} and \textsc{TeleoR}. It has discussed the implementation issues associated with several of the features of those two systems. Concepts behind teleo-reactive programming were discussed, such as the triple and two tower architectures.\\

A teleo-reactive system was built in Python, modelled on \textsc{TR} and \textsc{TeleoR}. To demonstrate its effectiveness, it was used to control a simple video game, throug the Pedro communications architecture. The implementation of the teleo-reactive system was discussed, with regards to some of the design decisions made, such as those concerning typed expressions. Example programs were provided.

\section*{Future Work}
If I had more time, I would have liked to implement concurrency in my application - Qulog provides one method but several other approaches have been discussed by others\\

To really demonstrate the utility of teleo-reactive programming, a more complex problem than ``Asteroids'' should be found. The problem should require a solution that is too complex for a single programmer to think about it in its entirety all at once. The ``Asteroids'' example was too simple, because the problem was small enough that a programmer could imagine the complete solution (programming an AI that wins the game by shooting all the asteroids) as one strategy. That said, ``Asteroids'' was a useful tool for debugging the Python implementation - I ensured that the simulation itself was stable by controlling it with Qulog, then replaced Qulog with my implementation.\\

Another aspect of teleo-reactive programming that should be better demonstrated is its robustness. A real-world demonstration would demonstrate this because of the complexity and unpredictable nature of the real world, compared to virtual worlds.  A good place to start would be with the Robot Operating System (ROS), which is a modular collection of robotics software (nodes) and a communications protocol to connect those nodes together. It supports publish/subscribe messaging using ``topics'' that nodes can subscribe to and direct peer-to-peer communication using ``services''\citep{quigley2009ros}. It also has a simulator, Gazebo, which means that control code can be tested out on a simulated robot and then run on a real life robot with the same physical attributes\citep{koenig2004design}.\\

Another area that would have been interesting to explore is a possible connection between teleo-reactive and functional reactive programming. Both of them concern outputs that vary depending on continuous input. Functional reactive programming has been used for robotics \citep{hudak2003arrows}.

\bibliography{bibliography}

\end{document}